\newif
\ifconfver
\confvertrue        
\ifconfver
    \documentclass[10pt,journal,final,twoside]{IEEEtran}
\else
    \documentclass[11pt,draftcls,onecolumn]{IEEEtran}
\fi

\usepackage{xspace,empheq,fancybox,amssymb,amsfonts,graphicx,epstopdf,epsfig,syntonly,times,subfigure} 
\usepackage{caption}
\usepackage{psfrag,color,bm,array,tabularx}
\usepackage{cite,url,footnote,xspace,syntonly,algorithmic}
\usepackage{verbatim,multirow}
\usepackage[linesnumbered,ruled,vlined]{algorithm2e}
\usepackage{nomencl}
\makenomenclature

\usepackage{textcomp}
\usepackage{xcolor}
\usepackage{mathtools}
\usepackage{bbm, booktabs}
\usepackage{amsmath, amsthm, lipsum}       
\usepackage{ifthen,etoolbox}
\usepackage{array}
\newcolumntype{C}[1]{>{\centering\arraybackslash}p{#1}}
\newcolumntype{R}[1]{>{\raggedleft\arraybackslash}p{#1}}

\newcolumntype{M}[1]{>{\centering\arraybackslash}m{#1}}
\newcolumntype{N}{@{}m{0pt}@{}}

\DeclareMathOperator
*{\argmax}{\arg\max}

\def\BibTeX{{\rm B\kern-.05em{\sc i\kern-.025em b}\kern-.08em
		T\kern-.1667em\lower.7ex\hbox{E}\kern-.125emX}}

\input{mysymbol.sty}
\allowdisplaybreaks

\newcommand{\KB}{\color{black}{}}

\newcommand{\KBP}{\color{black}{}}
\newcommand{\KBU}{\color{black}{}}
\IEEEoverridecommandlockouts
\renewcommand{\nomgroup}[1]{%
  \ifthenelse{\equal{#1}{A}}{\item[\textbf{Indices}]\vspace{8pt}}{
  \ifthenelse{\equal{#1}{B}}{\vspace{10pt}\item[\textbf{Sets}]\vspace{5pt}}{
  \ifthenelse{\equal{#1}{C}}{\vspace{10pt}\item[\textbf{Variables}]\vspace{5pt}}{
  \ifthenelse{\equal{#1}{D}}{\vspace{10pt}\item[\textbf{Parameters}]\vspace{5pt}}{}}}}%
}

\begin{document}
	


\title{\bf\LARGE Degradation-Aware Pumping Control of Variable-Speed Pumped Storage via Residual Reinforcement Learning}
\author{
\IEEEauthorblockN{Kyung-bin Kwon, \IEEEmembership{Member,~IEEE}, SangWoo Park, \IEEEmembership{Member,~IEEE},  and Dam Kim, \IEEEmembership{Member,~IEEE}}

\thanks{\protect\rule{0pt}{3mm} \textit{All authors contributed equally to this work. (Corresponding author: Dam Kim)}}
\thanks{\protect\rule{0pt}{3mm}  K.~Kwon is with the Optimization and Control Group, Pacific Northwest National Laboratory, Richland, WA 99352, USA (e-mail: kyung-bin.kwon@pnnl.gov)

~S. Park is with the Department of Mechanical and Industrial Engineering, New Jersey Institute of Technology, Newark, NJ 07103, USA (e-mail: sangwoo.park@njit.edu)

~D. Kim is with the Department of Convergence System Engineering, Chungnam National University, Daejeon 34134, South Korea (e-mail: damkim@cnu.ac.kr)
}} 

\markboth{IEEE Transactions on Power Systems (In Preparation)}%
{ \MakeLowercase{\textit{et al.}}: }
\renewcommand{\thepage}{}
\maketitle
\pagenumbering{arabic}

\begin{abstract}
Variable-speed pumped storage hydropower (VS-PSH) must honor short-block dispatch commitments while limiting the operational degradation that intensified regulation duty inflicts on its components. When a single controller pursues both aims at once, every tracking gain is paid for in degradation, a conflict that persists even under full model knowledge and look-ahead. This paper proposes a two-layer control architecture that separates the guaranteed commitment from the bounded learning. A deterministic feedforward-PI gate controller, auditable and certifiable for grid-connected operation, secures average power delivery over each five-minute block, while a residual reinforcement learning policy adjusts only the rotor speed within a fixed bound the gate loop can always absorb, so the worst-case command is bounded by construction. The speed policy tracks a demand-dependent best-efficiency-point reference and is trained against an operation-degradation index that combines off-best-efficiency hydraulic loss with power and actuation variation into one physically interpretable signal. {\KBU Across normal and stressed dispatch, the proposed policy lowers best-efficiency-point tracking error by roughly 96\% relative to a fixed-speed baseline and cuts total degradation by up to about 56\% under the most demanding dispatch. It matches or slightly exceeds a full-information model-based optimizer in efficiency while preserving substantially tighter block tracking.}
\end{abstract}



\section{Introduction}\label{sec:IN}
The accelerating integration of variable renewable energy is reshaping the operating environment of modern power systems. Deepening net-load variability intensifies the demand for fast, dispatchable flexibility~\cite{lannoye2012evaluation,zhao2015review}. Pumped storage hydropower (PSH) provides more than 90\% of global electricity storage volume and remains the backbone resource for absorbing this variability~\cite{iha2025world}. Flexible resources are coupled to system operations through market dispatch. Units participating in energy and balancing markets must deliver contracted power over short settlement blocks, often five minutes, and balancing-market penalties are typically assessed on block-average rather than instantaneous deviations~\cite{ferc2016settlement}. Faithful dispatch compliance is therefore not only a control objective but a market obligation. At the same time, a growing body of evidence shows that the intensified regulation duty imposed by renewable-rich systems accelerates wear, fatigue, and efficiency loss in hydropower units, which raises maintenance costs and shortens overhaul intervals~\cite{yang2016wear,yang2018burden}. 
Thus, how a flexibility resource executes its dispatch, and not only whether it does, becomes an operational question of direct relevance to power system operations.

Variable-speed pumped storage hydropower (VS-PSH) is particularly well positioned to address this tension. In conventional fixed-speed units, the pumping power is essentially dictated by the prevailing head, which leaves no continuous control authority and confines the machine to constant-load pumping~\cite{brown2008optimization,filipe2019optimal}. By decoupling the rotor speed from the grid frequency through power-electronic conversion, VS-PSH enables continuous pumping-power modulation~\cite{chen2023modeling}. Large-scale variable-speed plants are now in operation across Asia~\cite{guo2025variable} and Europe~\cite{filipe2019optimal}. Yet the speed degree of freedom carries a further, largely unexploited value. Because the pump's best-efficiency-point (BEP) speed varies with the demanded power level, a unit clamped at rated speed is hydraulically misaligned at virtually every operating point except one. Sustained off-BEP operation lowers hydraulic efficiency and narrows the usable operating range, while the fast, gate-dominated power corrections it necessitates induce pressure transients in the waterway~\cite{schmidt2017modeling}. Exploiting the speed degree of freedom for degradation management, rather than for power modulation alone, therefore represents a concrete opportunity to reconcile dispatch compliance with asset preservation.

\begin{table*}[!t]
\caption{Positioning Against Representative Prior Approaches}
\label{tab:positioning}
\centering
\setlength{\tabcolsep}{4pt}
\small
\begin{tabular}{lcccc}
\toprule
Method & Dispatch compliance & Speed DOF (BEP) & Degradation-aware & Bounded learning \\
\midrule
Scheduling \cite{chazarra2018optimal,liu2021secured,jukic2024optimal} & Block-level only & --- & No & N/A \\
Dynamic MPC \cite{jukic2023optimal} & Objective & Yes (eff.) & No & N/A \\
Wear filters \cite{yang2017wear} & --- & No & Yes (wear) & N/A \\
Stress MPC \cite{cassano2022stress} & Objective & No & Yes (fatigue) & N/A \\
RL grid ops \cite{yan2020multi,duan2020deep,chen2022reinforcement} & No & No & No & No \\
Safe RL \cite{wang2020safe} & Soft (in exp.) & No & No & Soft \\
Degradation-in-reward \cite{kabir2023deep} & No & No & Yes (reward) & No \\
Residual volt-var \cite{liu2025residual} & No & No & No & For opt. \\
Proposed & Yes (5-min, cert.) & Yes & Yes (ODI) & Yes (worst-case) \\
\bottomrule
\multicolumn{5}{p{0.98\textwidth}}{\footnotesize \vspace{2pt}
\textit{Dispatch compliance:} ``Yes'' = hard guarantee on the five-minute block;
``Objective'' = tracked within the controller's objective but not guaranteed;
``Soft'' = satisfied in expectation; ``Block-level only'' = resolved to the schedule block, not within it.
``---'' = not applicable to the method's setting; ``N/A'' = method is not learning-based.
Abbreviations: BEP = best-efficiency point; DOF = degree of freedom; ODI = operation-degradation index;
eff. = efficiency; exp. = expectation; opt. = optimization.} \\
\end{tabular}
\end{table*}

VS-PSH participation in electricity markets has been addressed primarily at the day-ahead and intra-day scheduling level~\cite{chazarra2018optimal,liu2021secured,jukic2024optimal}. Mixed-integer programming formulations co-optimize energy and reserve offers while capturing head-dependent capability limits, discrete mode transitions, and state-of-charge dynamics~\cite{chazarra2018optimal}. Related frameworks further address reserve security~\cite{liu2021secured} and multi-market participation~\cite{jukic2024optimal} for pumped storage. These scheduling layers, however, operate at hourly to sub-hourly granularity and resolve power commitments only down to the market settlement block, not the actuation within it. The second-by-second actuation that realizes each block is delegated to plant-level controllers, under the implicit assumption that execution is faithful and degradation-free. The intra-block dynamics remain outside the scheduling model. The model does not capture how the gate and speed actuators move, how far the machine strays from its BEP, or how much fatigue each block accumulates. Bridging this gap requires a real-time execution layer that honors the commitments issued from above while explicitly managing the degradation consequences of how those commitments are met.

At the plant-control level, prior work has only partially filled this role. Model predictive control (MPC) formulations capture the efficiency-optimal speed and gate settings of variable-speed units under pressure and actuation constraints, extending static optimization to closed-loop dynamic operation~\cite{jukic2023optimal}. However, these approaches optimize for efficiency and tracking alone, providing no feedback on accumulated degradation state. A complementary thread explicitly targets wear mitigation during regulation service. Controller dead-bands and filters reduce guide-vane movement at a quantified cost in regulation quality~\cite{yang2017wear}, and a stress-informed MPC scheme enforces the penstock fatigue limit through explicit stress constraints~\cite{cassano2022stress}. These approaches share a common structural limitation. Because both operate on conventional guide-vane-governed units whose rotor speed is locked to the grid, corrective authority is confined to the guide vanes, and neither exploits the speed degree of freedom. When tracking and degradation objectives are handled within a single optimization layer, the controller must continuously arbitrate between them, and improving one systematically sacrifices the other. As our benchmark results confirm, this structural trade-off persists even with full model knowledge and look-ahead.

Reinforcement learning (RL) has become an attractive approach for power system operations where accurate models are unavailable or costly to maintain. It has been applied to load-frequency control~\cite{yan2020multi} and autonomous voltage regulation~\cite{duan2020deep}, and more broadly across power system operations~\cite{chen2022reinforcement}. These studies show that learned policies can rival model-based controllers in average performance while acting far faster at execution time. Safety-critical dispatch, however, demands more than strong average performance. A learned policy gives no inherent guarantee that hard operating constraints will hold in the worst case, and a single violation during real-time execution can be operationally unacceptable. Safe reinforcement learning addresses this concern by embedding the constraints into the learning problem, for instance by formulating a constrained Markov decision process solved with a constrained actor-critic~\cite{wang2020safe}. Such formulations improve constraint satisfaction in expectation, yet they enforce the constraints softly and do not certify feasibility in the worst case.
Residual reinforcement learning offers a structurally different route to the same end. Rather than entrusting the entire control action to a learned policy, it adds a learned correction on top of a fixed base controller whose behavior is already understood, so that the base secures the essential objective while the residual improves performance~\cite{johannink2019residual}. The paradigm has recently entered power applications, where a residual policy refines an approximate model-based volt-var controller~\cite{liu2025residual}. In that work the residual action space is bounded to ease learning and sharpen optimization, not to certify the closed-loop behavior of the combined controller. The distinction is decisive for dispatch. If the base controller alone guarantees the hard commitment and the residual stays within a bound the base can always absorb, the worst-case behavior of the composite is certifiable by construction, independent of the policy that learning converges to.

Table~\ref{tab:positioning} positions the present work against representative prior approaches and exposes two gaps that none of them closes together. On the learning side, reinforcement learning policies for power system operations attain strong average performance but do not guarantee dispatch compliance in the worst case~\cite{yan2020multi,duan2020deep,chen2022reinforcement,wang2020safe}, and work that embeds a degradation term in the reward~\cite{kabir2023deep} still addresses a resource without a speed degree of freedom and leaves dispatch uncertified. On the degradation side, controllers that limit guide-vane movement or constrain penstock stress to a fatigue limit act on machines whose rotor speed is locked to the grid~\cite{yang2017wear,cassano2022stress}, so the speed degree of freedom central to VS-PSH stays unused.

This work closes both gaps with a two-layer architecture that keeps the certifiable commitment separate from the bounded learning, and its contributions are fourfold. First, the architecture pairs a deterministic feedforward proportional-integral (FF-PI) gate controller that guarantees compliance with each five-minute dispatch block with a residual proximal policy optimization (PPO) speed policy~\cite{schulman2017proximal} that operates within a fixed bound the gate controller can always absorb, so the worst-case command is bounded by construction. Second, the residual policy tracks the demand-dependent best-efficiency-point rather than holding the rated speed, removing the structural mismatch that a fixed-speed unit incurs at every demand level but one. Third, we formulate an operation-degradation index (ODI) that combines off-BEP hydraulic loss with power and actuation variation into a single composite of physically interpretable fatigue drivers, and embed it in the reward through the residual layer rather than the tracking loop, which substantially lowers degradation at only a modest, bounded cost in tracking accuracy. Fourth, we characterize the resulting trade-off between block tracking and degradation across normal and stressed dispatch, and show that the residual policy attains efficiency close to a full-information model-based optimizer while preserving substantially tighter tracking.

The remainder of this paper is organized as follows. Section~\ref{sec:formulation} develops the discrete-time VS-PSH pumping model, the demand-dependent BEP map, and the operation-degradation index, and casts dispatch execution as a block-tracking control problem. Section~\ref{sec:control} presents the two-layer architecture, in which a deterministic FF-PI gate controller secures block compliance and a residual PPO speed policy manages degradation within a fixed bound. Section~\ref{sec:num} reports numerical studies under normal and stressed dispatch, comparing the proposed controller against a fixed-speed baseline and a full-information model-based optimizer, and Section~\ref{sec:conclusion} concludes.

%

\section{VS-PSH Modeling and Problem Formulation}\label{sec:formulation}
\subsection{Discrete-Time Plant Model}

\begin{figure}[t]
\centering
\includegraphics[width=0.95\linewidth]{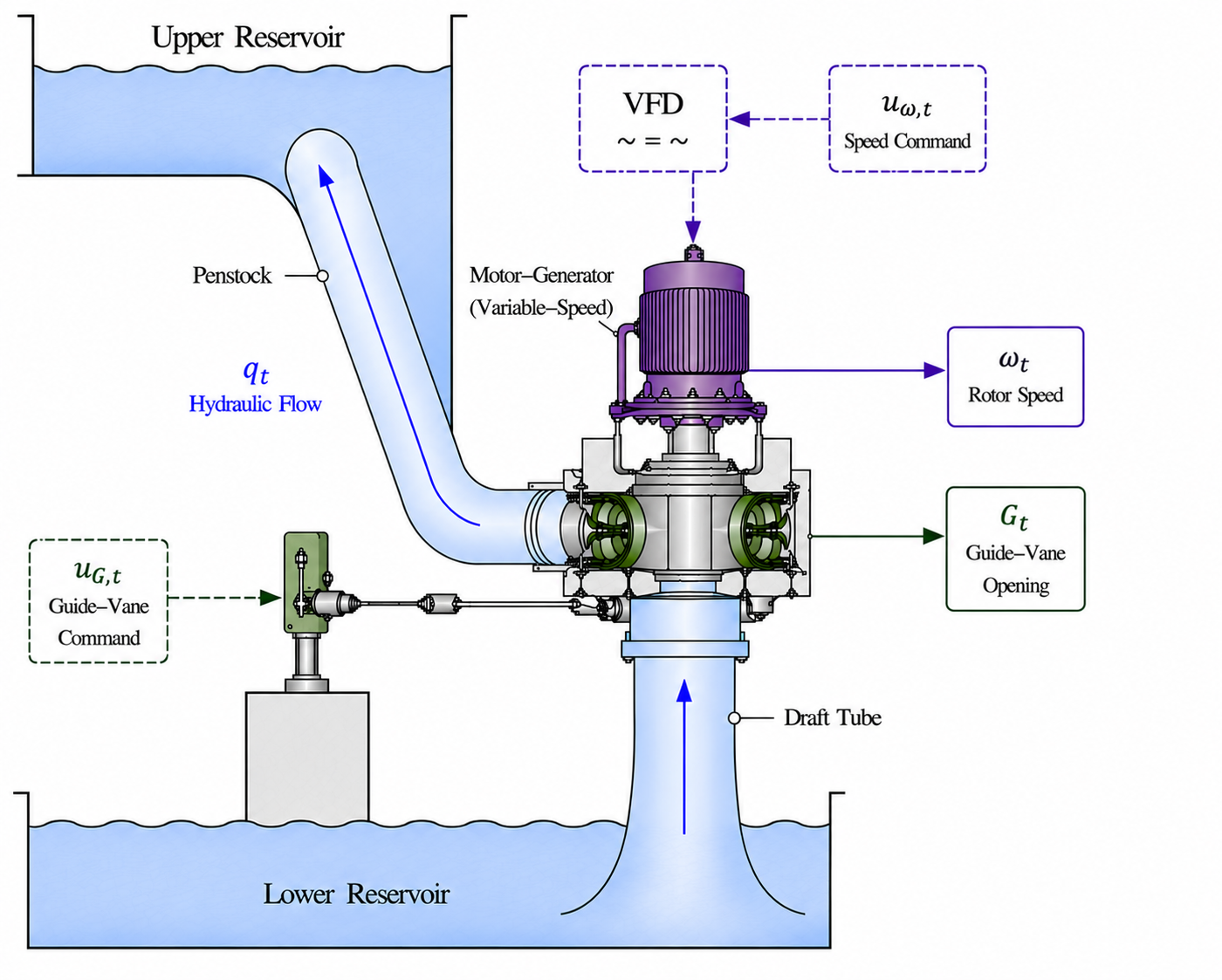}
\caption{VS-PSH pumping-mode actuation and plant variables}
\label{fig:vsps_schematic}
\end{figure}

We consider pumping-mode VS-PSH dynamics represented at 1-s resolution. Fig.~\ref{fig:vsps_schematic} summarizes the physical actuation paths and plant variables used in the discrete-time model. The hydraulic {\KBP flow rate} $q_t$, guide-vane opening $G_t$, and rotor speed $\omega_t$ are each modeled as first-order lag systems driven by their respective set-point commands:
\begin{subequations}
\begin{align}
q_{t+1} &= q_t + \frac{\Delta t}{\tau_q}\left(q_t^\star - q_t\right), \label{eq:flow_dyn} \\
G_{t+1} &= G_t + \frac{\Delta t}{\tau_G}\left(u_{G,t}-G_t\right), \label{eq:gate_dyn} \\
\omega_{t+1} &= \omega_t + \frac{\Delta t}{\tau_\omega}\left(u_{\omega,t}-\omega_t\right), \label{eq:speed_dyn}
\end{align}
\end{subequations}
where $\tau_q$, $\tau_G$, and $\tau_\omega$ are the hydraulic, gate actuator, and drive system time constants, respectively. The first-order structure captures the dominant inertial and hydraulic lag characteristics seen in VS-PSH commissioning data and remains computationally tractable for both analysis and control synthesis. The guide-vane command $u_{G,t}$ is the primary power-governing actuator; the speed command $u_{\omega,t}$ is issued by the variable-frequency drive to the motor-generator. 

The steady-state flow set-point $q_t^\star$ follows the pump affinity law \cite{schmidt2017modeling}:
\begin{align}
q_t^\star = Q_{\max}G_t\left(\frac{\omega_t}{\omega_r}\right), \label{eq:flow_target}
\end{align}
where $Q_{\max}$ is the maximum flow at fully-open guide vane and $\omega_r$ is the rated speed, reflecting the hydraulic similarity relation under constant head. The same affinity law gives the BEP flow at a given rotor speed as $q_{\mathrm{bep}}(\omega_t)=q_r(\omega_t/\omega_r)$.

The guide-vane ramp constraint
\begin{align}
|u_{G,t}-u_{G,t-1}| \le r_G\Delta t\label{eq:ramp}
\end{align}
enforces actuator rate limits arising from mechanical wear and hydraulic transient considerations; $r_G$ is a hard constraint enforced throughout all simulations. {\KBP Both $G_t$ and $u_{G,t}$ are normalized guide-vane opening variables in per-unit, with 0 denoting fully closed and 1 denoting fully open.}

The instantaneous pumping power {\KBP in MW scale} is
\begin{align}
P_t={\rho g H q_t \eta_t}, \label{eq:power}
\end{align}
where $\rho$ is water density, $g$ gravitational acceleration, and $H$ the effective hydraulic head \cite{yang2019advantage}. Power can be maintained by increasing $q_t$ at lower $\eta_t$ (off-BEP, gate-dominated) or by adjusting $\omega_t$ to improve $\eta_t$. The latter motivates the residual speed control layer.

\subsection{BEP Map and Efficiency-Dependent Degradation Proxy}
The BEP speed is a function of the demanded power level $d_t$. In VS-PSH units, the pump-turbine specific speed is fixed by the runner geometry, so that for a given head and flow demand there exists a unique rotor speed that maximizes hydraulic efficiency. This speed-demand relationship is approximated by the affine BEP map \cite{jukic2023optimal}
{\KBP
\begin{align}
\omega_{\mathrm{bep}}(d_t) = \left[\omega_r - k_{\mathrm{bep}}
\frac{d_t-d_{\mathrm{mid}}}{d_{\mathrm{half}}}
\right]_{\omega_r-k_{\mathrm{bep}}}^{\omega_r+k_{\mathrm{bep}}}
\label{eq:bep_map} 
\end{align}
}
where {\KBP $[x]_a^b \triangleq \min(\max(x, a), b)$,} $k_{\mathrm{bep}}$ is the maximum speed deviation from rated, $d_{\mathrm{mid}}$ is the midpoint of the operating demand range, and $d_{\mathrm{half}}$ is the half-range normalizing factor. The clipping operation in \eqref{eq:bep_map} enforces drive system speed limits and ensures that the BEP reference remains physically realizable. For this machine the map produces an \emph{inverse} speed-demand relationship: at lighter loading the pump attains its efficiency optimum at a speed \emph{above} rated ($\omega_{\mathrm{bep}}>{\omega_r}$ when $d_t < d_{\mathrm{mid}}$), while at heavier loading the BEP speed falls below rated; at $d_t = d_{\mathrm{mid}}$ the two coincide. FF-PI's constant rated speed is therefore sub-optimal at every demand level except $d_{\mathrm{mid}}$.

Following recent VS-PSH efficiency-aware modeling practice, hydraulic efficiency is represented by a quadratic penalty model around the operating BEP \cite{schmidt2017modeling, chen2023modeling}:
\begin{align}
\eta_t = \eta_{\max}\Big(
1
-w_q \xi_{q,t}^2
-w_\omega \xi_{\omega,t}^2
-w_h \xi_{h,t}^2
-w_g \ell_g(G_t)
\Big), \label{eq:eta}
\end{align}
where $\eta_{\max}$ is the peak hydraulic efficiency, and the normalized deviations from BEP conditions are
\begin{align}
&\xi_{q,t}=\frac{q_t-q_{\mathrm{bep}}(\omega_t)}{q_r},\quad
\xi_{\omega,t}=\frac{\omega_t-\omega_{\mathrm{bep}}(d_t)}{\omega_n}, \nonumber \\ 
&\xi_{h,t}=\frac{h_t-h_r}{h_r}, q_{\mathrm{bep}}(\omega_t)=q_r\!\left(\frac{\omega_t}{\omega_r}\right), \nonumber \\
&\ell_g(G_t)=\max(0,G_t-0.92)^2+\max(0,0.08-G_t)^2, \label{eq:lg}
\end{align}
with $q_r$ the rated flow, $\omega_n$ the speed normalizing factor, and $h_r$ the rated head. Each squared penalty in \eqref{eq:eta} captures a distinct hydraulic loss: $\xi_{q,t}$ penalizes recirculation losses from off-BEP flow; $\xi_{\omega,t}$ captures the speed-demand mismatch that directly motivates the residual speed policy; $\xi_{h,t}$ accounts for off-design head conditions (a constant offset under fixed-head operation, retained for generality); and $\ell_g(G_t)$ penalizes extreme guide-vane positions. The weights $w_q$, $w_\omega$, and $w_g$ are identified from manufacturer hill charts.

The normalized efficiency shortfall below the BEP reference is captured by the off-BEP loss:
\begin{align}
\ell_{\mathrm{off},t}=\max\!\left(0,1-\frac{\eta_t}{\eta_{\max}}\right). \label{eq:loff}
\end{align}
This term is zero when $\eta_t = \eta_{\max}$ (perfect BEP operation) and increases proportionally to the fractional efficiency drop, and it serves as a direct measure of hydraulic degradation potential.

Building on $\ell_{\mathrm{off},t}$, the ODI is defined as a weighted sum of three physically interpretable fatigue drivers:
\begin{align}
\mathrm{ODI}_t =
w_P \frac{|P_t-P_{t-1}|}{P_{\mathrm{ref}}}
+w_G \frac{|G_t-G_{t-1}|}{r_G}
+w_{\mathrm{off}} \ell_{\mathrm{off},t}, \label{eq:odi}
\end{align}
where the three terms represent the normalized power fluctuation rate, guide-vane actuation rate, and off-BEP hydraulic loss, respectively, with weights $w_P$, $w_G$, and $w_{\mathrm{off}}$. Each captures a distinct fatigue driver: power fluctuation stresses motor-generator windings; rapid guide-vane movement induces servo-hydraulic wear and pressure transients; off-BEP operation promotes cavitation and runner erosion. Normalizing by $P_{\mathrm{ref}}$ and $r_G$ ensures dimensional homogeneity.

The ODI is accumulated into a smoothed degradation state through an exponential moving average (EMA):
\begin{align}
D_{t+1}=(1-\alpha_D)D_t+\alpha_D\mathrm{ODI}_t, \label{eq:ema}
\end{align}
where $\alpha_D$ is the forgetting factor. $D_t$ provides the policy with a smoothed signal of recent operating severity, included in the observation vector $x_t$ so the policy can modulate speed to reduce long-term degradation.

\subsection{Tracking Objective and Control Problem}
VS-PSH units participating in frequency regulation or energy markets are typically dispatched with 5-minute block commitments, meaning the unit must deliver a contracted average power $d_k$ over each 300-second interval. The block-average power and its tracking error are
\begin{align}
\bar{P}_k=\frac{1}{N}\sum_{t=kN}^{kN+N-1}P_t,\qquad
e_{\mathrm{blk},k}=\bar{P}_k-d_k, \label{eq:blk_err}
\end{align}
where $N$ is the number of 1-second samples per block \cite{chazarra2018optimal}. The block error $e_{\mathrm{blk},k}$ is the primary grid-compliance metric. Penalties in balancing markets are typically assessed on block-average deviations, not instantaneous ones.

To enable intra-block feedback before the final settlement, two causal online tracking errors are defined within each active block:
\begin{align}
e_{\mathrm{run},t}=\bar{P}_{k,t}^{\mathrm{partial}}-d_k,\qquad
e_{\mathrm{req},t}=P_t-\bar{P}_{k,t}^{\mathrm{req}}, \label{eq:online_err}
\end{align}
{\KBP where $\bar{P}_{k,t}^{\mathrm{partial}}$ is the running average from the beginning of block $k$ to the current time. The required remaining average is computed as
\begin{align}
\bar{P}_{k,t}^{\mathrm{req}} = \frac{N d_k-\sum_{i=kN}^{t-1}P_i}{N-(t-kN)}, \label{eq:req_power}
\end{align}
for $t=kN,\ldots,kN+N-1$.} This is the average power that must be delivered over the remaining samples of the block, including the current sample, in order to satisfy $\bar{P}_k=d_k$ at the block boundary. Together, these signals give a richer picture of block progress than a simple instantaneous error.

The overall control objective combines intra-block online errors, block-boundary compliance, and the degradation state into a single expected-cost minimization:
\begin{align}
\min_{\pi}\ \mathbb{E}\Bigg[
&\sum_t
\left(
\lambda_{\mathrm{run}}(e_{\mathrm{run},t})^2
+\lambda_{\mathrm{req}}(e_{\mathrm{req},t})^2
+\lambda_{\mathrm{odi}}\mathrm{ODI}_t
\right)
\nonumber \\
&+\sum_k \lambda_{\mathrm{blk}}(e_{\mathrm{blk},k})^2
\Bigg].
\label{eq:obj}
\end{align}
The inner sum balances step-level power delivery against degradation; the outer sum enforces block-boundary compliance. 

%
\section{Degradation-Aware Control Architecture}\label{sec:control}
\subsection{Two-Layer Structure}
The proposed architecture separates control into two layers. The lower layer is the FF-PI gate controller that maps block-tracking error to a guide-vane command, ensuring $\bar{P}_k \approx d_k$ even under model uncertainty. {\KBP The controller is deterministic and does not rely on learned components, {\KBU which facilitates} auditing, verification, and certification for grid-connected applications.}

\begin{figure}[t]
\centering
\includegraphics[width=\linewidth]{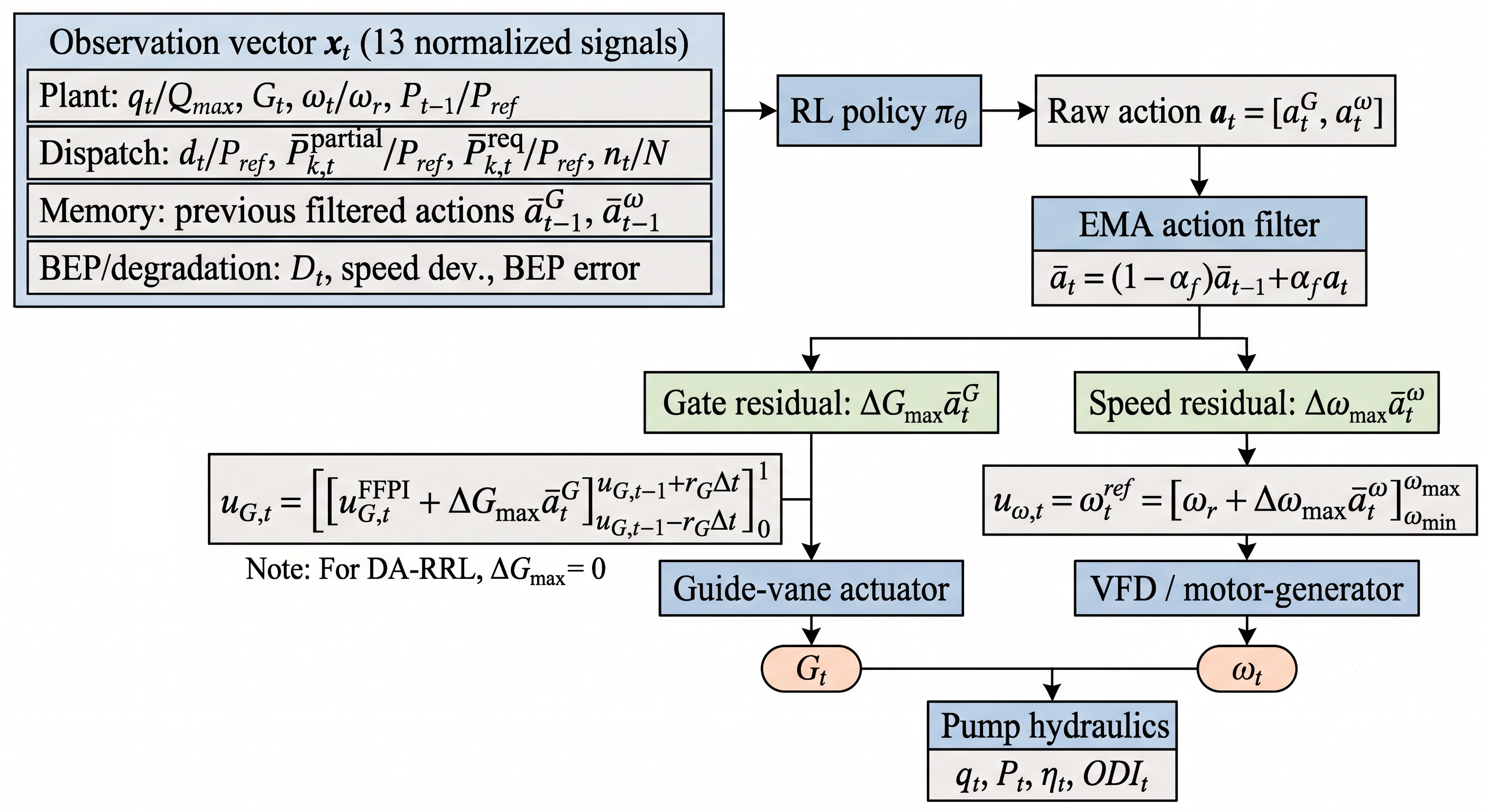}
\caption{Residual-RL signal flow from observation vector to plant commands.}
\label{fig:rl_flowchart}
\end{figure}

{\KB
The upper layer learns a bounded residual correction to the actuator commands, as summarized in Fig.~\ref{fig:rl_flowchart}. The RL policy produces the raw residual action
\begin{align}
a_t = [a_t^G,\ a_t^\omega]^\top \in [-1,1]^2,
\end{align}
where $a_t^G$ is the optional guide-vane residual component and $a_t^\omega$ is the speed residual component. To suppress abrupt command changes, the raw action is passed through an exponential moving average,
\begin{align}
\bar{a}_t=(1-\alpha_f)\bar{a}_{t-1}+\alpha_f a_t, \label{eq:action_filter}
\end{align}
where $\alpha_f$ is the filter coefficient. The filtered speed component is converted into the rotor-speed reference
\begin{align}
u_{\omega,t}=\omega_t^{\mathrm{ref}}=\left[\omega_r+\Delta\omega_{\max}\bar{a}_t^\omega
\right]_{\omega_{\min}}^{\omega_{\max}}, \label{eq:residual}
\end{align}
which is the speed command applied to the drive dynamics in \eqref{eq:speed_dyn}. When a guide-vane residual is enabled, the plant guide-vane command first enforces the ramp-rate limit \eqref{eq:ramp} on the candidate command and then saturates the result to the physical opening range,
\begin{align}
u_{G,t}=\left[
\left[u_{G,t}^{\mathrm{FFPI}}+\Delta G_{\max}\bar{a}_t^G\right]_{u_{G,t-1}-r_G\Delta t}^{u_{G,t-1}+r_G\Delta t}
\right]_0^1. \label{eq:gate_cmd}
\end{align}

As shown in Fig.~\ref{fig:rl_flowchart}, the 13-dimensional observation vector $x_t$ is grouped into four categories. The plant-state group contains $q_t/Q_{\max}$, $G_t$, $\omega_t/\omega_r$, and $P_{t-1}/P_{\mathrm{ref}}$. The dispatch and block-compliance group contains $d_t/P_{\mathrm{ref}}$, $\bar{P}_{k,t}^{\mathrm{partial}}/P_{\mathrm{ref}}$, $\bar{P}_{k,t}^{\mathrm{req}}/P_{\mathrm{ref}}$, and $n_t/N$. The memory/action group contains the previous filtered residual actions $\bar{a}_{t-1}^G$ and $\bar{a}_{t-1}^\omega$. The BEP/degradation group contains $D_t$, the normalized speed deviation from rated, and the normalized BEP tracking error. This grouping gives the policy visibility into plant state, block-compliance status, recent control memory, and degradation-relevant BEP alignment.
}
The four controllers compared in this study are:
\begin{itemize}
\item \textbf{FF-PI}: The fixed-speed baseline in which $\omega_t = \omega_r$ throughout, and power tracking is achieved entirely by gate modulation. This controller represents current industrial practice for fixed-speed pumped storage and serves as the primary tracking reference.
\item \textbf{\KBP Tracking-Prioritized Residual RL (TP-RRL)}: FF-PI augmented with a residual RL policy trained using a tracking-only reward ($\lambda_{\mathrm{odi}}=\lambda_\eta=0$). The policy adjusts both speed (primary) and gate (small residual offset) as a residual correction over FF-PI. This variant serves as a tracking-prioritized no-degradation baseline: its reward weights are calibrated to emphasize block tracking and end-of-block correction in the absence of any degradation-related learning signal.
\item \textbf{\KBP Degradation-Aware Residual RL (DA-RRL)}: FF-PI augmented with a residual RL policy trained using the full degradation-aware reward. This is the proposed method.
\item \textbf{MPC-D}: A receding-horizon MPC benchmark that optimizes both tracking and degradation objectives over a finite look-ahead horizon, solved at each time step by grid search. MPC-D represents the performance ceiling of a single-layer model-based optimizer that has full access to the degradation model; its limitations reveal the inherent trade-off between tracking and degradation when both objectives are handled without structural decoupling.
\end{itemize}

\subsection{Reward and Learning Design}
The per-step reward for both RL methods is
\begin{align}
r_t =&-\lambda_{\mathrm{blk}}\tilde{e}_{\mathrm{blk},k,t}^{2}
-\lambda_{\mathrm{run}}\tilde{e}_{\mathrm{run},t}^{2}
-\lambda_{\mathrm{req}}\tilde{e}_{\mathrm{req},t}^{2}
-\lambda_{\mathrm{odi}}\mathrm{ODI}_t \nonumber \\
&-\lambda_{\omega}\!\left(\frac{\omega_t-\omega_r}{{\KBU \Delta\omega_{\max}}}\right)^{\!2}
-\lambda_{\mathrm{ctrl}}\|\bar{a}_t\|_1
-\lambda_{\mathrm{jerk}}\|\bar{a}_t-\bar{a}_{t-1}\|_1 \nonumber\\
&+\lambda_{\eta}\frac{\eta_t}{\eta_{\max}}, \label{eq:reward}
\end{align}
where $\tilde{e}$ terms are tracking errors normalized by $P_{\mathrm{ref}}$, and $\bar{a}_t \in [-1,1]$ is the filtered policy output (see {\KBU \eqref{eq:action_filter}}) \cite{schulman2017proximal}. The quadratic tracking penalties ($\lambda_{\mathrm{blk}}$, $\lambda_{\mathrm{run}}$, $\lambda_{\mathrm{req}}$) cover multiple timescales; $\lambda_{\mathrm{odi}}\mathrm{ODI}_t$ incentivizes BEP operation; $\lambda_\omega$ penalizes speed deviation from rated (active in TP-RRL to prevent unnecessary speed excursions); $\lambda_{\mathrm{ctrl}}\|\bar{a}_t\|_1$ and $\lambda_{\mathrm{jerk}}\|\bar{a}_t-\bar{a}_{t-1}\|_1$ suppress oscillations; and $\lambda_\eta\eta_t/\eta_{\max}$ provides a dense efficiency bonus that stabilizes early training. For DA-RRL, a tracking deadband is applied so that block, running, and required-power penalties are zero within specified tolerances, allowing the RL policy to concentrate on speed optimization once tracking is sufficiently accurate.

{\KBP Table~\ref{tab:reward_cfg} lists the reward weights. {\KBP The different tracking-related weights reflect controller-specific calibration rather than an attempt to favor either method.} TP-RRL uses $\lambda_{\mathrm{blk}} = 1200$ with $\lambda_{\mathrm{odi}} = \lambda_\eta = 0$, concentrating effort entirely on tracking. Since no degradation-related learning signal is available in TP-RRL, larger intra-block and required-power penalties are used to make it a tracking-prioritized baseline and to enforce block-end correction. This higher $\lambda_{\mathrm{req}}$ can also explain the sharper end-of-block behavior observed in the time-domain trajectories. In contrast, DA-RRL raises $\lambda_{\mathrm{blk}}$ to $1600$, reduces intra-block weights (trusting FF-PI for fine correction), and activates the ODI penalty and efficiency bonus; therefore, its intra-block tracking weights are relaxed, and a tracking deadband is used so that the upper-layer RL policy can exploit the speed degree of freedom for BEP adherence and degradation reduction. PPO is used for its {\KBU stable} on-policy updates with dense reward signals \cite{schulman2017proximal}.}

\subsection{MPC-D Benchmark}
To provide a model-based constrained benchmark, the MPC-D controller solves at each time step the finite-horizon problem:
\begin{align}
\min_{\{u_{G,\tau},u_{\omega,\tau}\}_{\tau=t}^{t+H-1}}
\Bigg[
&\sum_{\tau=t}^{t+H-1}\!\!\left(
w_{\mathrm{run}}(e_{\mathrm{run},\tau})^2
+w_{\mathrm{odi}}\mathrm{ODI}_\tau
\right)
\nonumber \\
&+w_{\mathrm{term}}(e_{\mathrm{blk},k,\mathrm{end}})^2
\Bigg],
\label{eq:mpc}
\end{align}
subject to the predicted plant dynamics, actuator bounds, and guide-vane ramp constraint
\begin{align}
&0 \le u_{G,\tau}^{\mathrm{cmd}} \le 1,\quad
|u_{G,\tau}^{\mathrm{cmd}}-u_{G,\tau-1}^{\mathrm{cmd}}| \le r_G\Delta t,
\nonumber\\
&\omega_{\min}\le u_{\omega,\tau}\le\omega_{\max}.
\end{align}
The predicted states are propagated using the same first-order guide-vane, rotor-speed, and flow dynamics as in the evaluation simulator. In implementation, the optimization is approximated through grid search over $N_G$ gate candidates and $N_\omega$ speed candidates within a horizon of $H$ seconds, with weights $w_{\mathrm{run}}$, $w_{\mathrm{term}}$, $w_{\mathrm{odi}}$. Infeasible guide-vane changes are clipped by the ramp limiter before evaluating the rollout cost. The terminal cost penalizes predicted block-average error, providing look-ahead for compliance. This benchmark exposes the tracking-degradation conflict that arises when both objectives are handled within a single optimization layer.

\begin{algorithm}[t]
\caption{Degradation-Aware Residual VS-PSH Control}
\KwIn{Demand block sequence $\{d_k\}$, FF-PI parameters, trained policy $\pi_\theta$}
\KwOut{Gate command $u_{G,t}$ and speed reference $u_{\omega,t}$}
Initialize plant state $(q_0,G_0,\omega_0)$ and degradation state $D_0$\;
\For{$t=0,1,\dots$}{
Update FF-PI baseline gate target from current block-tracking error\;
Construct state vector $x_t$ including $\bar{P}_{k,t}^{\mathrm{partial}}$, $\bar{P}_{k,t}^{\mathrm{req}}$, $D_t$, and BEP error\;
{\KBU Compute raw residual action $a_t=\pi_\theta(x_t)$\;}
{\KBU Update EMA filtered action $\bar{a}_t = (1-\alpha_f)\bar{a}_{t-1} + \alpha_f a_t$\;}
{\KBU Set $u_{\omega,t}=\left[\omega_r+\Delta\omega_{\max}\bar{a}_t^\omega\right]_{\omega_{\min}}^{\omega_{\max}}$\;}
{\KBU Set $u_{G,t}=\left[\left[u_{G,t}^{\mathrm{FFPI}}+\Delta G_{\max}\bar{a}_t^G\right]_{u_{G,t-1}-r_G\Delta t}^{u_{G,t-1}+r_G\Delta t}\right]_0^1$ (where $\Delta G_{\max}=0$ for DA-RRL)\;}
Apply gate rate limits and update $(q_{t+1},G_{t+1},\omega_{t+1})$\;
Compute $\eta_t$, $\mathrm{ODI}_t$, and $D_{t+1}$\;
At block boundary, update FF trim using block-average tracking error\;
}
\end{algorithm}

%

\section{Numerical Studies}\label{sec:num}
\subsection{Simulation Setup}

\begin{table}[t]
\caption{Plant, Model, and Controller Parameters}
\label{tab:plant_params}
\centering
\footnotesize
\setlength{\tabcolsep}{3pt}
\renewcommand{\arraystretch}{1.1}
\begingroup
\hyphenpenalty=10000
\exhyphenpenalty=10000
\begin{tabularx}{\columnwidth}{@{}l >{\raggedright\arraybackslash}X r@{}}
\toprule
\multicolumn{3}{l}{\textit{Plant parameters}} \\
\midrule
$H$ & Hydraulic head & 100 m \\
$Q_{\max}$ & Maximum flow & 120 m$^3$/s \\
$q_r$ & Rated flow & 80 m$^3$/s \\
$\omega_r$ & Rated speed & 50 Hz \\
$\omega_{\min}/\omega_{\max}$ & Drive speed limits & 45/55 Hz \\
$P_{\mathrm{ref}}$ & Rated power reference & 80 MW \\
$\eta_{\max}$ & Peak efficiency & 0.92 -- \\
$G_t,u_{G,t}$ & Guide-vane opening/command range & 0--1 p.u. \\
$\tau_q,\tau_G,\tau_\omega$ & Time constants & 30/8/20 s \\
$r_G$ & Guide-vane ramp limit & 0.004 p.u.\,s$^{-1}$ \\
$N$ & Samples per 5-min block & 300 -- \\
\midrule
\multicolumn{3}{l}{\textit{Efficiency model parameters (hill charts)}} \\
\midrule
$w_q,w_\omega,w_g,w_h$ & Efficiency penalty weights & 0.20/0.15/0.10/0.03 -- \\
$h_r$ & Rated head & 95 m \\
$\omega_n$ & Speed normalizing factor & 5 Hz \\
\midrule
\multicolumn{3}{l}{\textit{BEP map parameters}} \\
\midrule
$k_{\mathrm{bep}}$ & Max BEP speed deviation & 3 Hz \\
$d_{\mathrm{mid}},d_{\mathrm{half}}$ & BEP map center/half-range & 50/20 MW \\
\midrule
\multicolumn{3}{l}{\textit{ODI and degradation state parameters}} \\
\midrule
$w_P,w_G,w_{\mathrm{off}}$ & ODI weights & 1.00/0.05/0.70 -- \\
$\alpha_D$ & EMA forgetting factor & 0.02 -- \\
\midrule
\multicolumn{3}{l}{\textit{Residual RL policy parameters}} \\
\midrule
$\Delta\omega_{\max}$ & Max speed deviation (DA-RRL/T) & 4/3 Hz \\
$\Delta G_{\max}$ & Max gate residual (DA-RRL/TP-RRL) & {\KBU 0/0.05 p.u.} \\
$\alpha_f$ & Action filter coefficient & 0.35 -- \\
Deadband (blk/run/req) & Tracking tolerances (DA-RRL) & 0.50/0.80/3.00 MW \\
\midrule
\multicolumn{3}{l}{\textit{MPC-D parameters}} \\
\midrule
$H$ & Prediction horizon & 45 s \\
$N_G,N_\omega$ & Grid-search candidates & 17/9 -- \\
$w_{\mathrm{run}},w_{\mathrm{term}},w_{\mathrm{odi}}$ & MPC weights & 15/3500/1.5 -- \\
\bottomrule
\end{tabularx}
\endgroup
\end{table}

All controllers are evaluated on a simulation horizon of 3600~s (60 minutes, i.e., 12 consecutive 5-min blocks) at 1-s control resolution. {\KBP The plant and model parameters are selected to represent a medium-head VS-PSH unit and are consistent with typical variable-speed pumped-storage modeling ranges reported in \cite{schmidt2017modeling,yang2019advantage,jukic2023optimal}. The resulting parameter set is summarized in Table~\ref{tab:plant_params}.}

Two demand schedules are designed to test different aspects of controller performance. The \textit{normal} scenario subjects the unit to stochastic moderate ramps with demand increments drawn uniformly from $[-8, 8]$~MW and clipped to the operating range $[30, 70]$~MW, representing typical intra-hour market dispatch signals. The \textit{stress} scenario imposes frequent high-variation demand changes following a pre-defined pattern with inter-block transitions of up to 15~MW, simulating the demanding load-following conditions that arise during periods of high renewable variability or system contingency events. The stress scenario is intentionally designed to expose the tension between rapid tracking and BEP adherence, which is the key challenge motivating the proposed architecture.

{\KBP RL policies are trained for 700,000 environment steps using PPO with a multilayer perceptron (MLP) policy, $n_{\mathrm{steps}}=2{,}048$, a batch size of 256, 10 epochs per update, $\gamma=0.99$, $\lambda_{\mathrm{GAE}}=0.95$, a learning rate of $3\times10^{-4}$, and a clip range of 0.2.} Evaluation is performed on 10 fixed random seeds held out from training. All four methods (FF-PI, MPC-D, TP-RRL, DA-RRL) share identical plant dynamics, actuator constraints ($r_G$, $\omega_{\min}$, $\omega_{\max}$), and ODI weight parameters, ensuring a fair comparison. {\KBP The evaluation metrics are the block-average tracking mean absolute error (MAE), measured in MW; the mean BEP-tracking error, $|\omega_t-\omega_{\mathrm{bep}}(d_t)|$, measured in Hz; the mean off-BEP ODI component, $\overline{w_{\mathrm{off}}\ell_{\mathrm{off}}}$; and the mean total ODI.}

\begin{table}[t]
\caption{Reward weights for learning-based controllers.}
\label{tab:reward_cfg}
\centering
\setlength{\tabcolsep}{4pt}
\small
{\KBU
\begin{tabular}{lcccccccc}
\toprule
Method & $\lambda_{\mathrm{blk}}$ & $\lambda_{\mathrm{run}}$ & $\lambda_{\mathrm{req}}$ & $\lambda_{\mathrm{odi}}$ & $\lambda_{\eta}$ & $\lambda_\omega$ & $\lambda_{\mathrm{ctrl}}$ & $\lambda_{\mathrm{jerk}}$ \\
\midrule
TP-RRL & 1200 & 10.0 & 5.0  & 0   & 0  & 2.0 & 0.003 & 0.0 \\
DA-RRL & 1600 & 1.5  & 0.15 & 40  & 12 & 0   & 0.02 & 0.10 \\
\bottomrule
\end{tabular}
}
\end{table}

\subsection{Main Quantitative Results}
Tables~\ref{tab:normal_metrics} and \ref{tab:stress_metrics} summarize the core performance metrics for all four controllers in the normal and stress scenarios, respectively.

\begin{table}[t]
\caption{Main Metrics in Normal Scenario}
\label{tab:normal_metrics}
\centering
\setlength{\tabcolsep}{5pt}
{\KBU
\begin{tabular}{lcccc}
\toprule
Method & MAE (MW) & BEP Err (Hz) & Off-BEP ODI & Total ODI \\
\midrule
FF-PI  & 0.191 & 1.268 & 0.0102 & 0.0134 \\
MPC-D  & 0.616 & 0.076 & 0.0002 & 0.0060 \\
TP-RRL & 0.218 & 1.258 & 0.0101 & 0.0157 \\
DA-RRL & 0.300 & 0.052 & 0.0001 & 0.0036 \\
\bottomrule
\end{tabular}
}
\end{table}

\begin{table}[t]
\caption{Main Metrics in Stress Scenario}
\label{tab:stress_metrics}
\centering
\setlength{\tabcolsep}{5pt}
{\KBU
\begin{tabular}{lcccc}
\toprule
Method & MAE (MW) & BEP Err (Hz) & Off-BEP ODI & Total ODI \\
\midrule
FF-PI  & 1.164 & 1.237 & 0.0092 & 0.0156 \\
MPC-D  & 1.295 & 0.169 & 0.0010 & 0.0120 \\
TP-RRL & 0.290 & 1.194 & 0.0089 & 0.0188 \\
DA-RRL & 1.179 & 0.060 & 0.0004 & 0.0068 \\
\bottomrule
\end{tabular}
}
\end{table}

The results reveal the following performance characteristics across methods. In the normal scenario (Table~\ref{tab:normal_metrics}), FF-PI achieves the best tracking MAE ({\KBU 0.191~MW}), but fixed-speed operation leaves a large BEP mismatch ({\KBU 1.268~Hz}). MPC-D improves BEP behavior and ODI, but with severe tracking degradation ({\KBU 0.616~MW MAE}). TP-RRL moderately improves BEP error ({\KBU 1.258~Hz}) but does not reduce total ODI versus FF-PI. DA-RRL gives the strongest overall degradation-side performance in normal operation: BEP error drops to {\KBU 0.052~Hz (about 96\%} lower than FF-PI) and total ODI to {\KBU 0.0036 (about 73\%} lower than FF-PI), while MAE increases to {\KBU 0.300~MW}.

The stress scenario results (Table~\ref{tab:stress_metrics}) show a sharper trade-off. TP-RRL achieves the smallest MAE ({\KBU 0.290~MW}), indicating very strong tracking under stress, but with limited ODI improvement. DA-RRL instead prioritizes degradation suppression: total ODI is the lowest among all methods ({\KBU 0.0068}), about {\KBU 56\%} lower than FF-PI and about {\KBU 43\%} lower than MPC-D, while maintaining much better tracking than MPC-D ({\KBU 1.179 vs.\ 1.295~MW} MAE). BEP error for DA-RRL ({\KBU 0.060~Hz}) is {\KBU even better than} MPC-D ({\KBU 0.169~Hz}), confirming effective BEP-side control.

\subsection{Time-Domain Performance Analysis}
\begin{figure}[t]
\centering
\subfigure[Normal-scenario block tracking comparison]{
\includegraphics[width=\columnwidth]{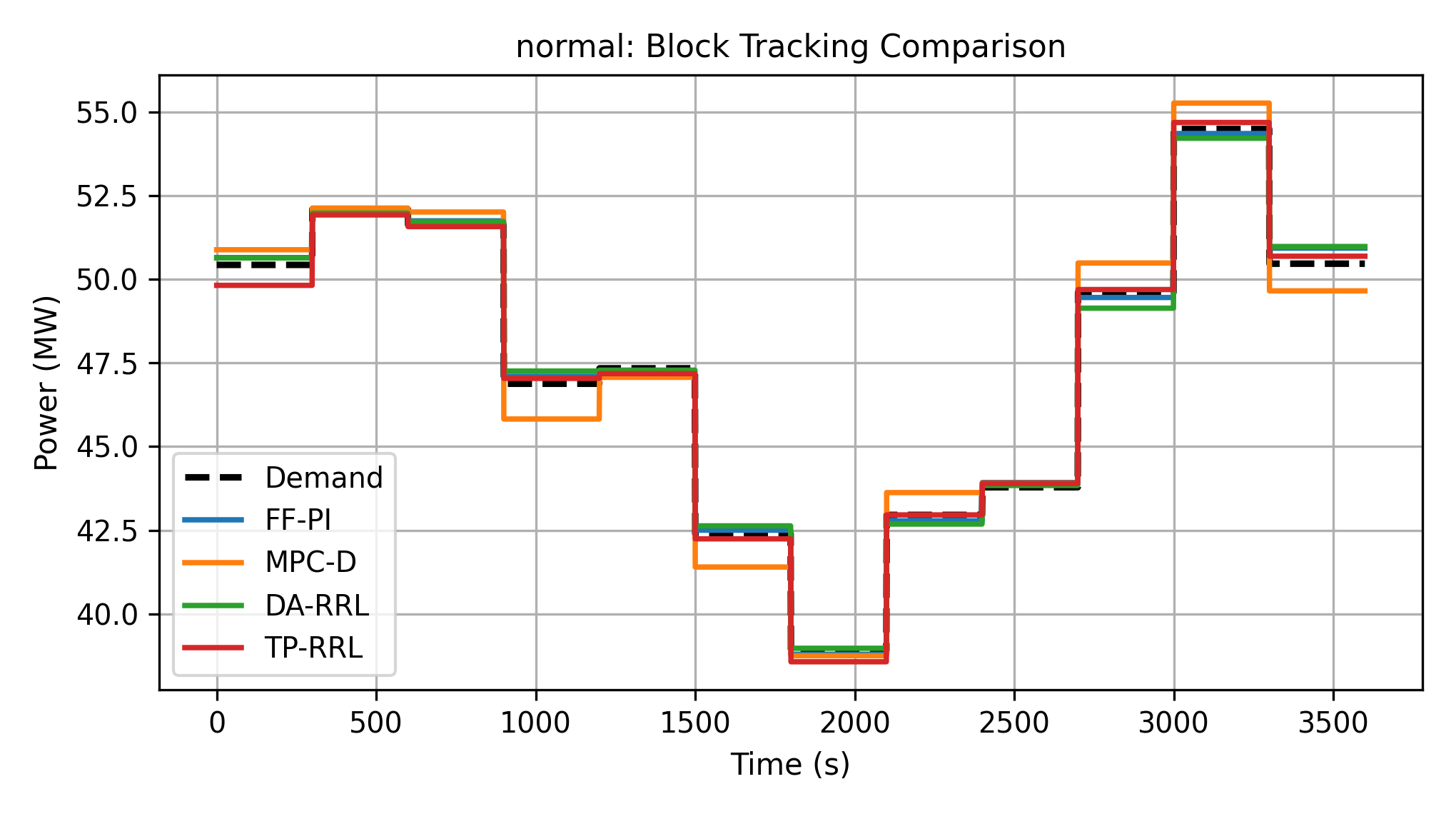}
}
\\[1mm]
\subfigure[Speed trajectory vs.\ BEP speed map]{
\includegraphics[width=\columnwidth]{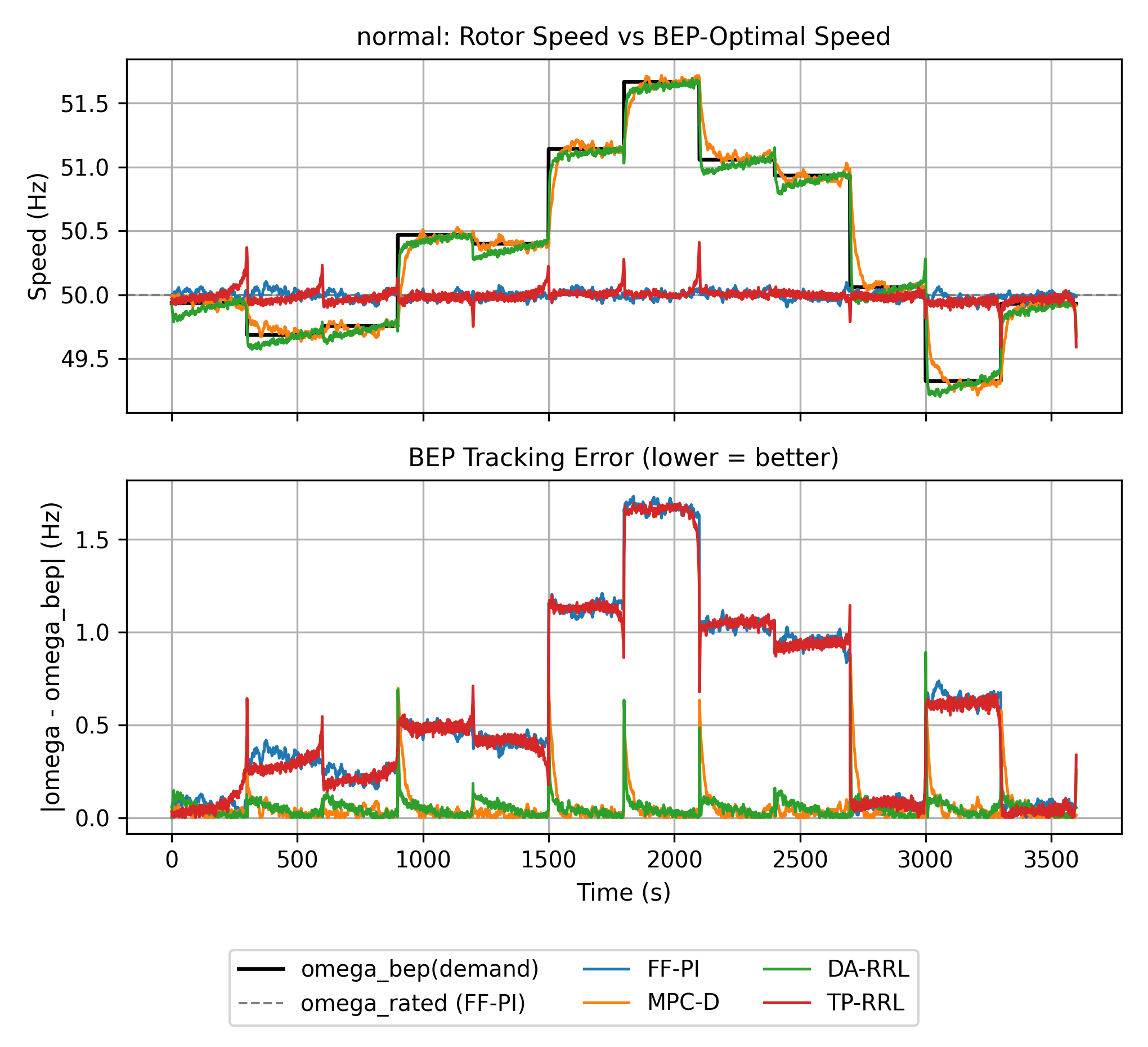}
}
\caption{Normal-scenario block tracking (a) and rotor speed relative to the demand-dependent BEP map (b).}
\label{fig:normal_tracking_bep}
\end{figure}

\begin{figure}[t]
\centering
\subfigure[Stress-scenario block tracking comparison]{
\includegraphics[width=\columnwidth]{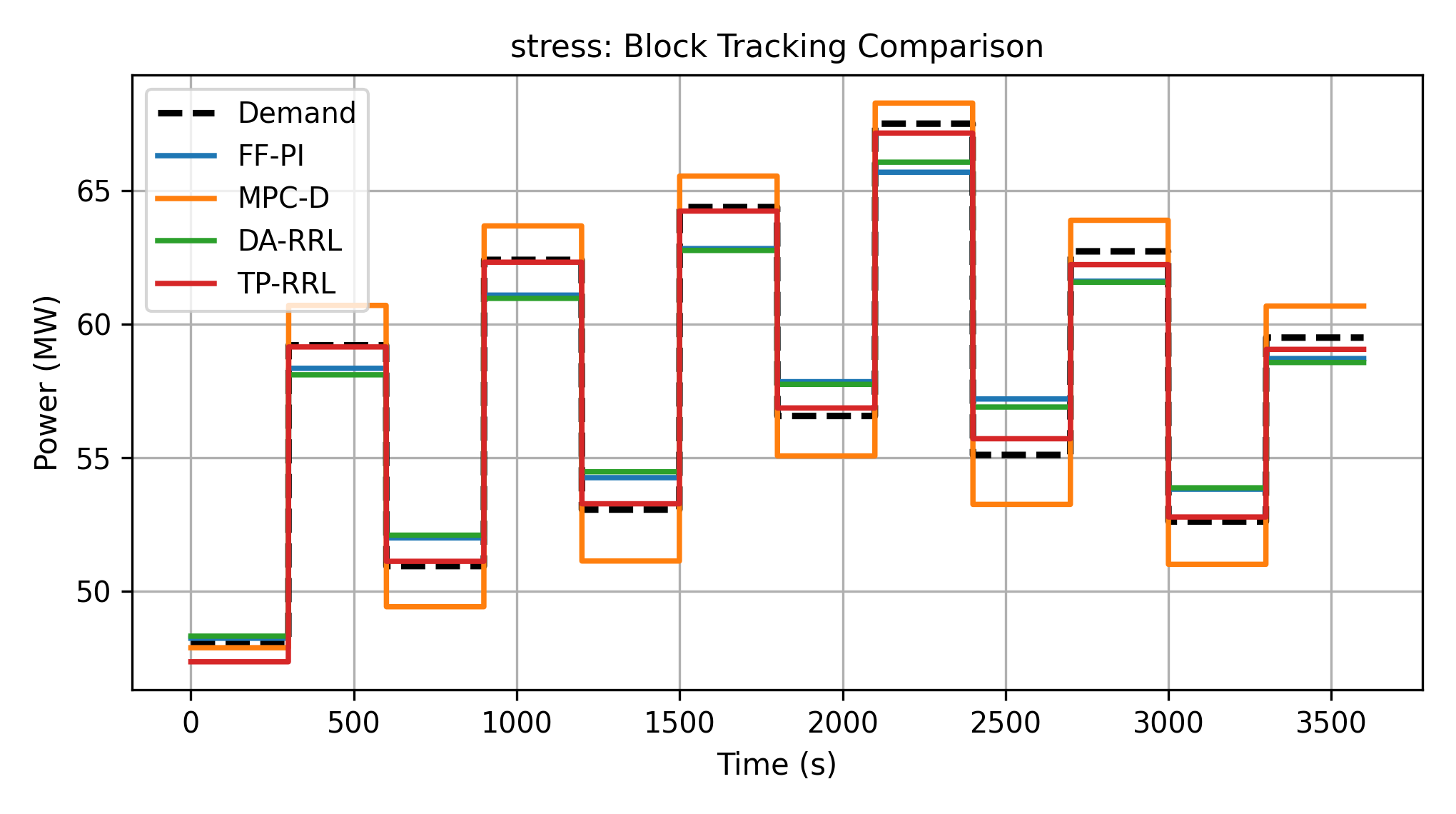}
}
\\[1mm]
\subfigure[Stress-scenario speed behavior vs.\ BEP map]{
\includegraphics[width=\columnwidth]{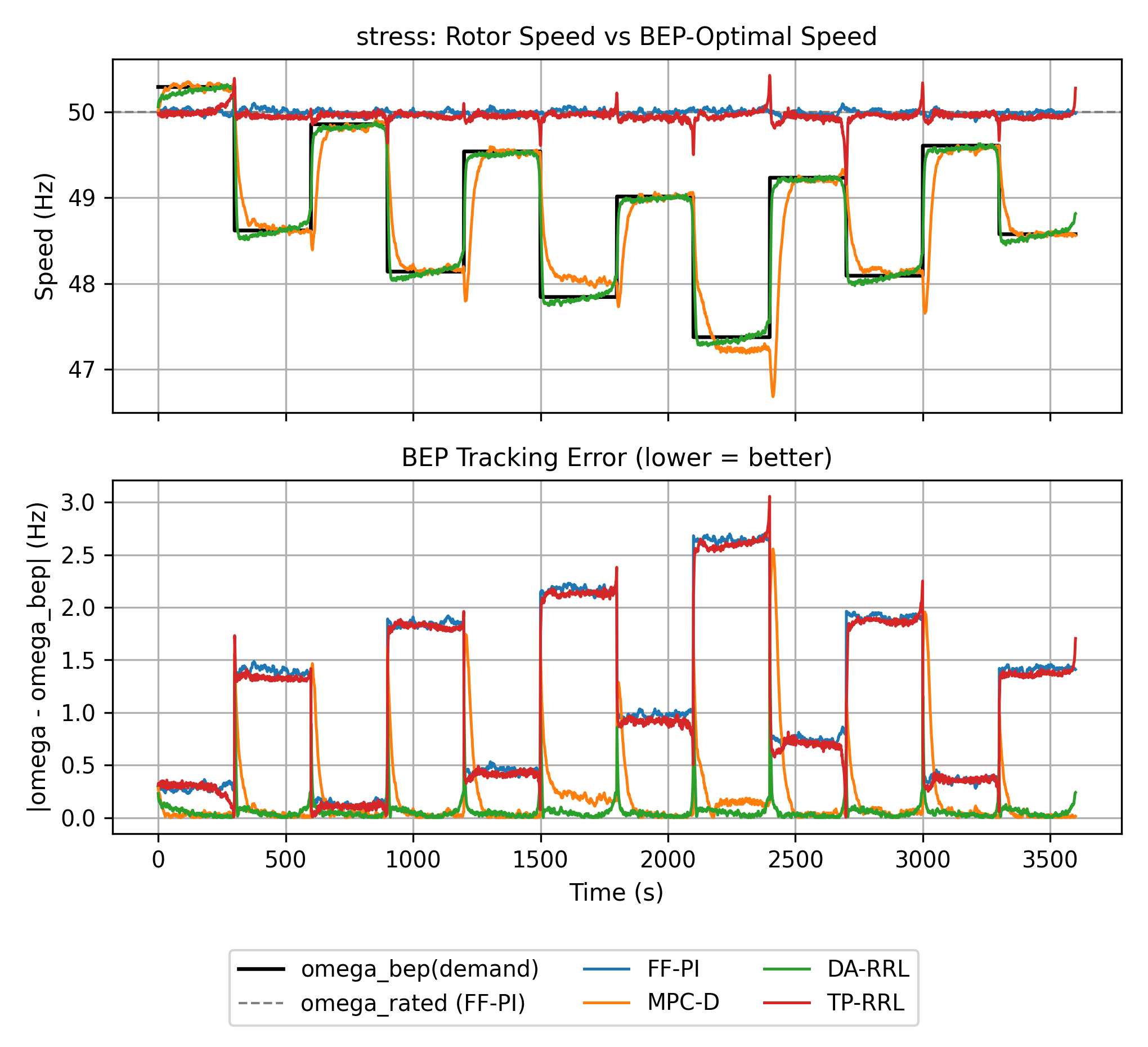}
}
\caption{Stress-scenario block tracking (a) and rotor speed relative to the demand-dependent BEP map (b).}
\label{fig:stress_tracking_bep}
\end{figure}

Figures~\ref{fig:normal_tracking_bep} and \ref{fig:stress_tracking_bep} compare power tracking and rotor speed against the BEP schedule for the normal and stress scenarios, respectively. In both cases, FF-PI and TP-RRL maintain tighter block power delivery, while DA-RRL and MPC-D follow $\omega_{\mathrm{bep}}(d_t)$ closely at the cost of larger power deviations. The speed panels make the BEP mismatch of FF-PI explicit: because the rotor runs at $\omega_r$ regardless of demand, the BEP error accumulates across the full load range, consistent with the 1.268~Hz (normal) and 1.237~Hz (stress) figures in Tables~\ref{tab:normal_metrics} and~\ref{tab:stress_metrics}. Under stress, TP-RRL achieves the tightest block tracking (0.215~MW MAE) but provides limited BEP improvement, whereas DA-RRL maintains a more balanced trade-off, matching MPC-D's BEP tracking while incurring only a fraction of MPC-D's tracking penalty.

\begin{figure}[t]
\centering
\subfigure[Normal-scenario metric bars]{
\includegraphics[width=\columnwidth]{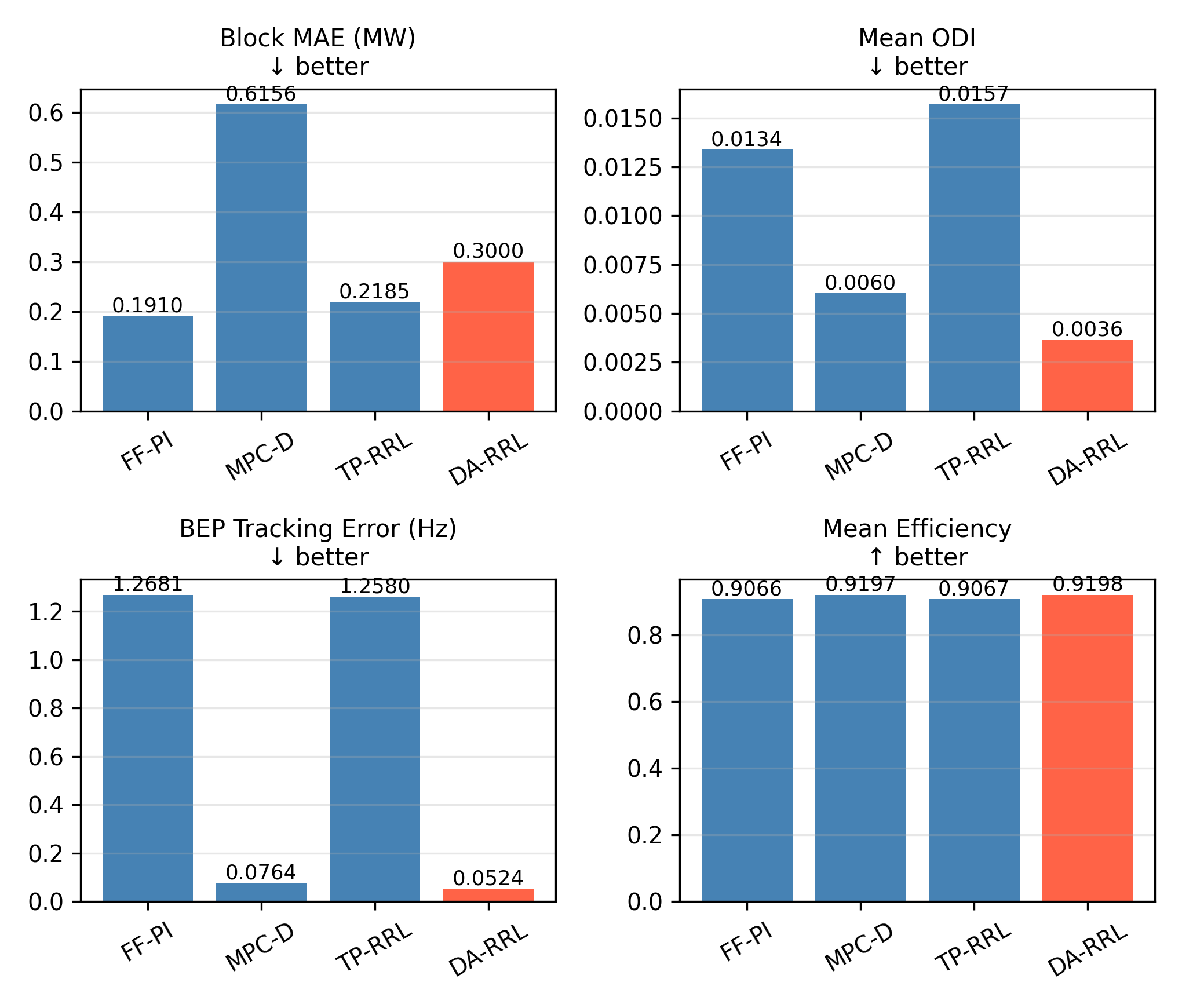}
}
\\[1mm]
\subfigure[Stress-scenario metric bars]{
\includegraphics[width=\columnwidth]{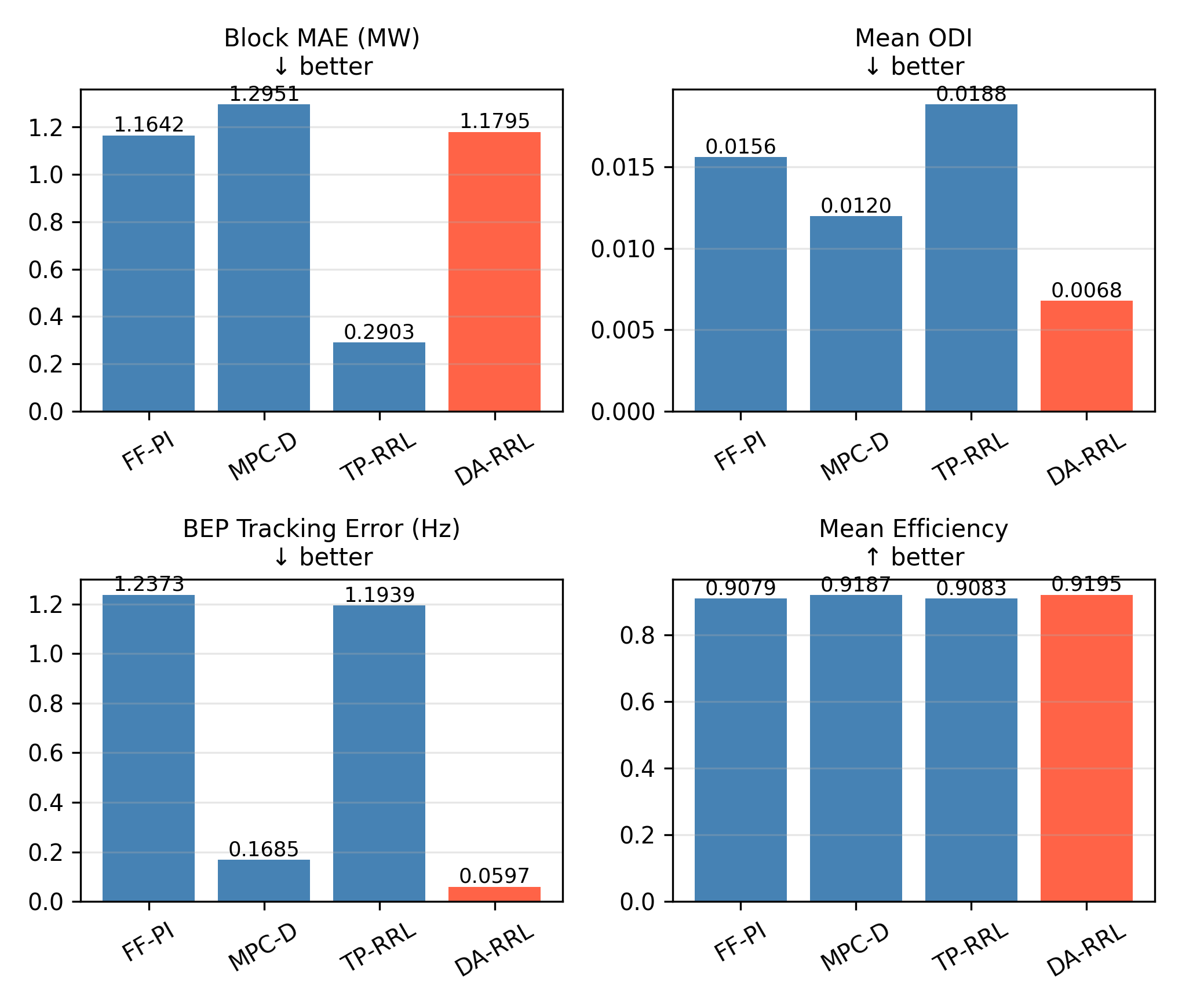}
}
\caption{Method-wise comparison of block MAE, total ODI, BEP tracking error, and mean efficiency in the normal (a) and stress (b) scenarios.}
\label{fig:metric_bars}
\end{figure}

Figure~\ref{fig:metric_bars} consolidates all four evaluation metrics across both scenarios in a single bar chart. The panels confirm that DA-RRL achieves the lowest BEP error and total ODI in both operating conditions, with mean efficiency closely approaching MPC-D's level, while MAE remains well below MPC-D in both scenarios.

\begin{figure}[t]
\centering
\subfigure[Stress-scenario ODI trajectory]{
\includegraphics[width=\columnwidth]{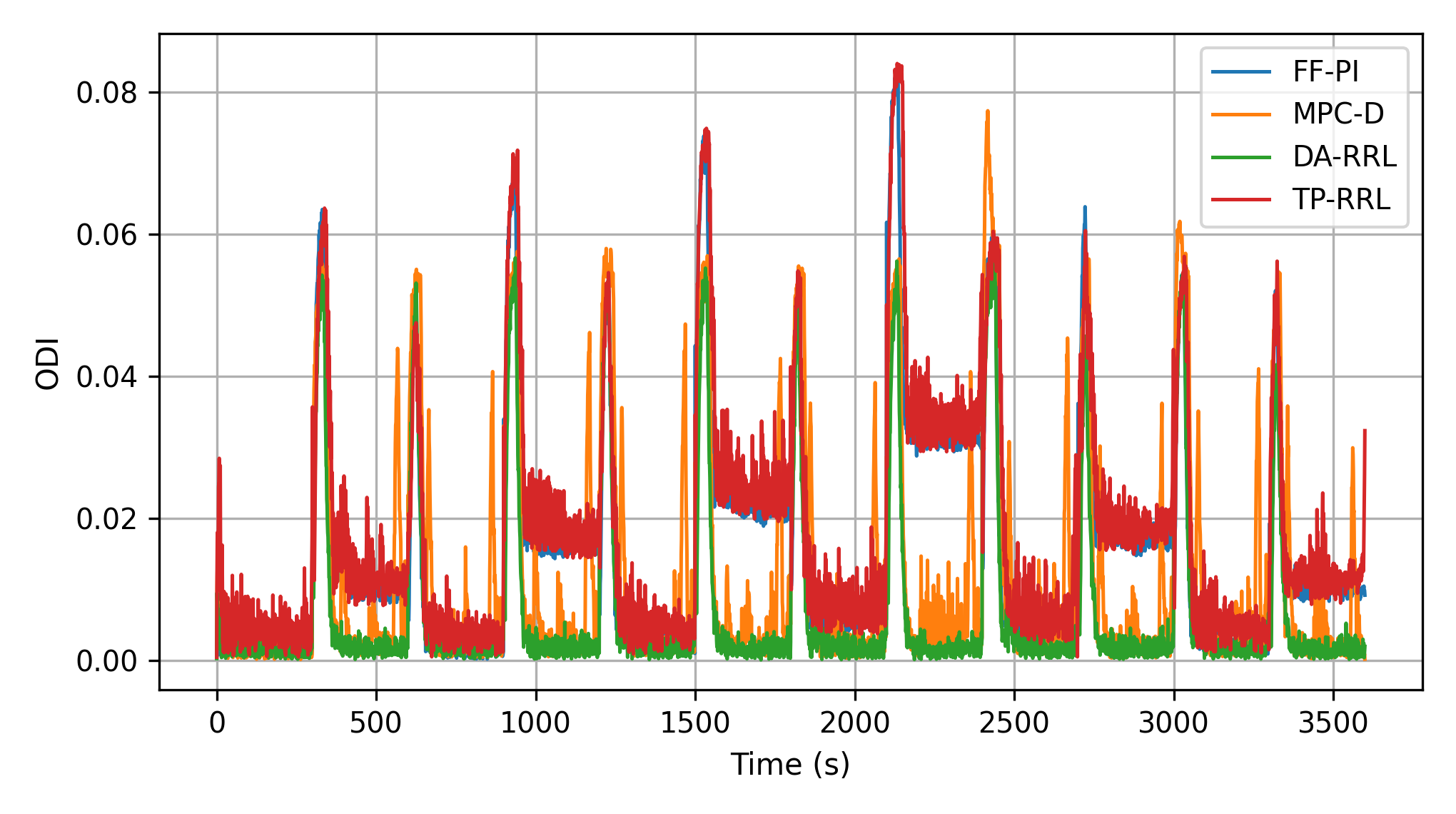}
}
\\[1mm]
\subfigure[Stress-scenario efficiency trajectory]{
\includegraphics[width=\columnwidth]{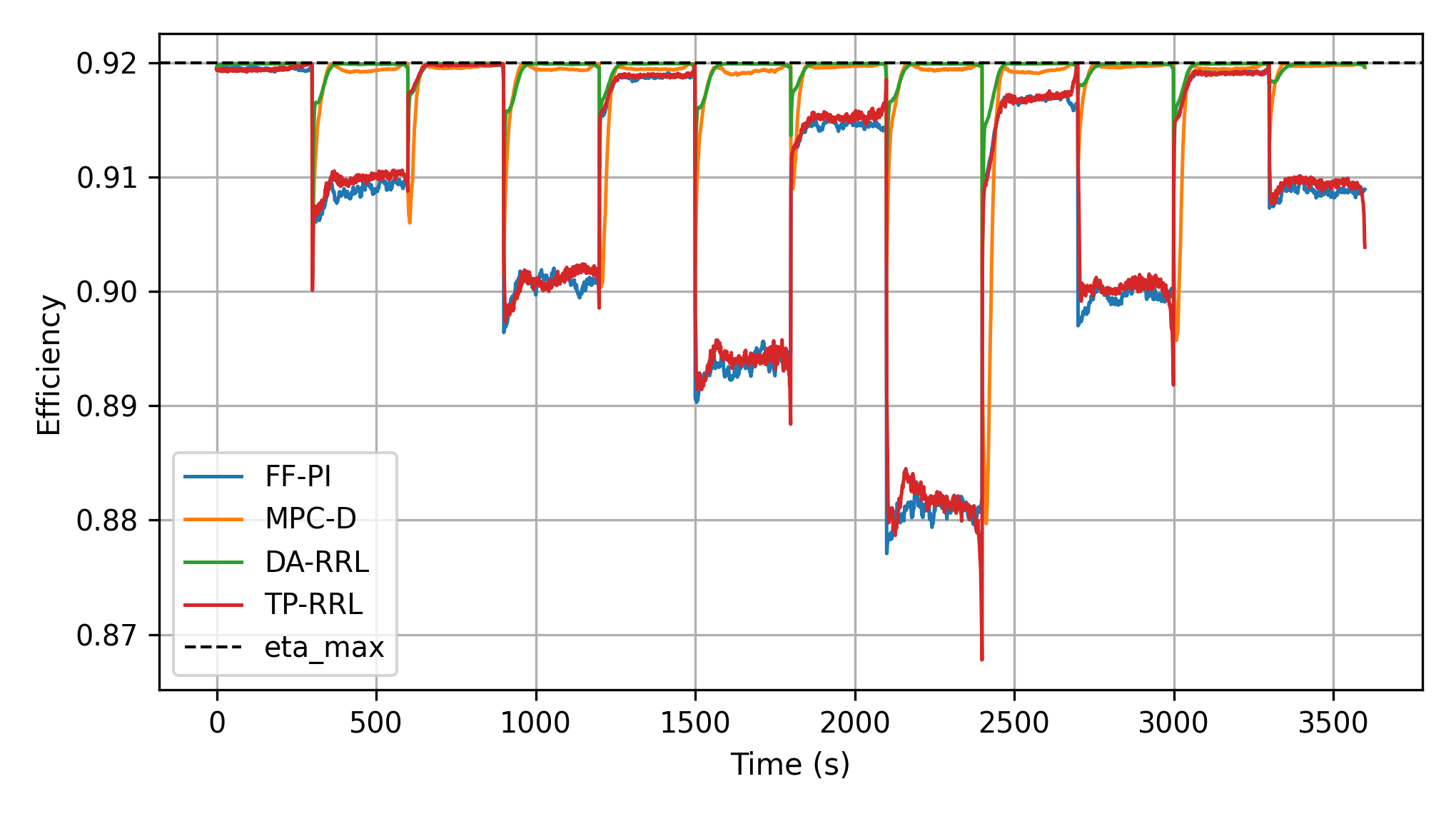}
}
\caption{Stress-scenario time trajectories of ODI (a) and hydraulic efficiency (b).}
\label{fig:stress_mechanism}
\end{figure}

\begin{figure}[t]
\centering
\includegraphics[width=\columnwidth]{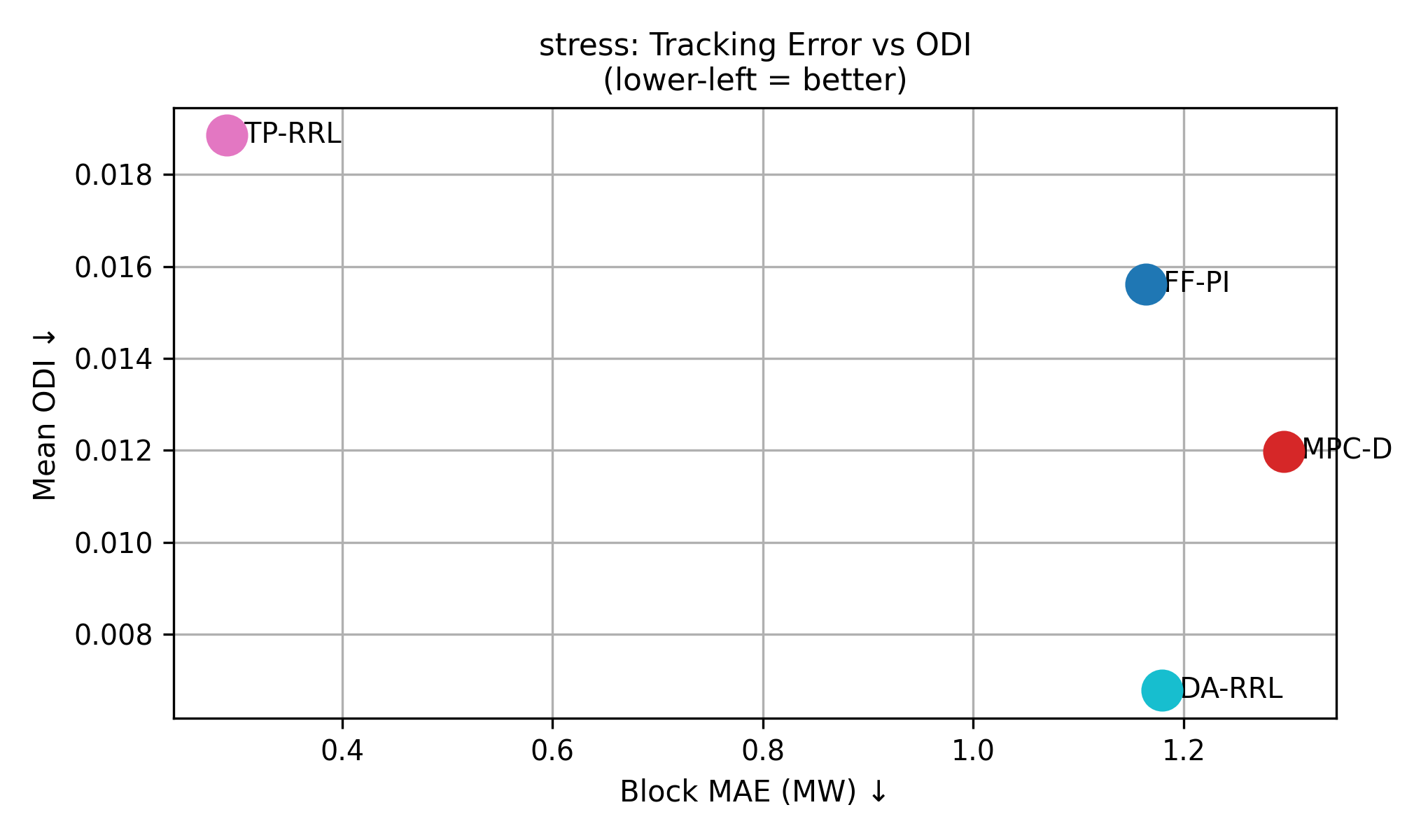}
\caption{Stress-scenario tracking–degradation trade-off in the block-MAE versus mean-ODI plane.}
\label{fig:pareto_stress_only}
\end{figure}

The stress-scenario time trajectories in Fig.~\ref{fig:stress_mechanism} provide mechanistic evidence for the ODI reduction. DA-RRL's instantaneous ODI trace follows MPC-D closely throughout, and its efficiency consistently exceeds that of FF-PI and TP-RRL. This confirms that the degradation benefit stems from sustained BEP adherence rather than from merely reduced actuation frequency. Figure~\ref{fig:pareto_stress_only} summarizes the overall tracking-degradation trade-off in the MAE-ODI plane: DA-RRL lies on the practical compromise frontier between FF-PI (tracking-prioritized) and MPC-D (ODI-prioritized), achieving the lowest total ODI of any method while preserving substantially better tracking than MPC-D.

\subsection{Sensitivity Analysis}
\begin{figure}[t]
\centering
\includegraphics[width=\columnwidth]{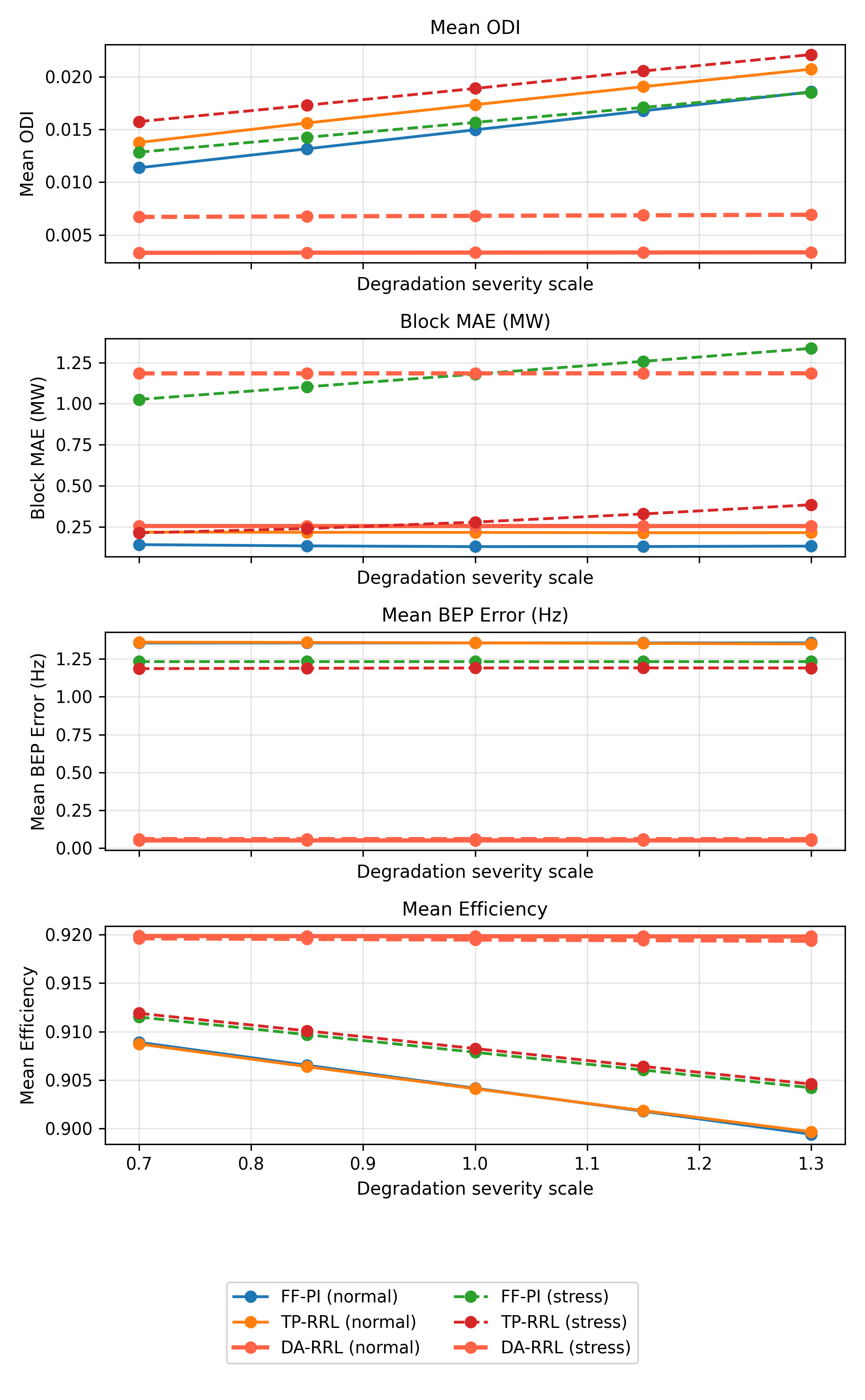}
\caption{Sensitivity to degradation-severity scale $s_d\in[0.7,1.3]$: mean ODI, block MAE, BEP error, and efficiency in normal and stress scenarios.}
\label{fig:sensitivity}
\end{figure}

Two sensitivity tests are conducted to assess robustness without retraining any controller. In the first test, the efficiency-penalty weights are scaled by a global degradation-severity factor $s_d\in[0.7,1.3]$, where $s_d<1$ corresponds to an optimistic model and $s_d>1$ to a pessimistic one. Figure~\ref{fig:sensitivity} plots the four evaluation metrics across $s_d$ for six fixed seeds per scenario, and Table~\ref{tab:sens_summary} reports the corresponding worst-case deviations from the nominal ($s_d=1.0$) values.

\begin{table}[t]
\caption{Sensitivity Summary (Worst-Case Change from $s_d=1.0$)}
\label{tab:sens_summary}
\centering
\begin{tabular}{llccc}
\toprule
Scenario & Method & $\Delta$ODI\textsubscript{max} & $\Delta$MAE\textsubscript{max} & $\Delta\eta$\textsubscript{drop} \\
\midrule
Normal & FF-PI      & +24.48\% & +27.38\% & 1.41\% \\
Normal & DA-RRL & +24.26\% & +7.34\%  & 1.10\% \\
Stress & FF-PI      & +19.74\% & +11.58\% & 0.59\% \\
Stress & DA-RRL & +12.68\% & +3.15\%  & 0.21\% \\
\bottomrule
\end{tabular}
\end{table}

DA-RRL shows considerably tighter robustness than FF-PI, particularly under stress: ODI growth is limited to $+12.68\%$ versus $+19.74\%$ for FF-PI, MAE growth to $+3.15\%$ versus $+11.58\%$, and efficiency drop to $0.21\%$ versus $0.59\%$. Across the full perturbation range, DA-RRL maintains an ODI advantage of 19.8--20.4\% over FF-PI in the normal scenario and 40.3--48.5\% under stress, with a consistent lead over TP-RRL of roughly 20\% (normal) and 48\% (stress).

\begin{figure}[t]
\centering
\subfigure[Normal-scenario ODI sensitivity]{
\includegraphics[width=\columnwidth]{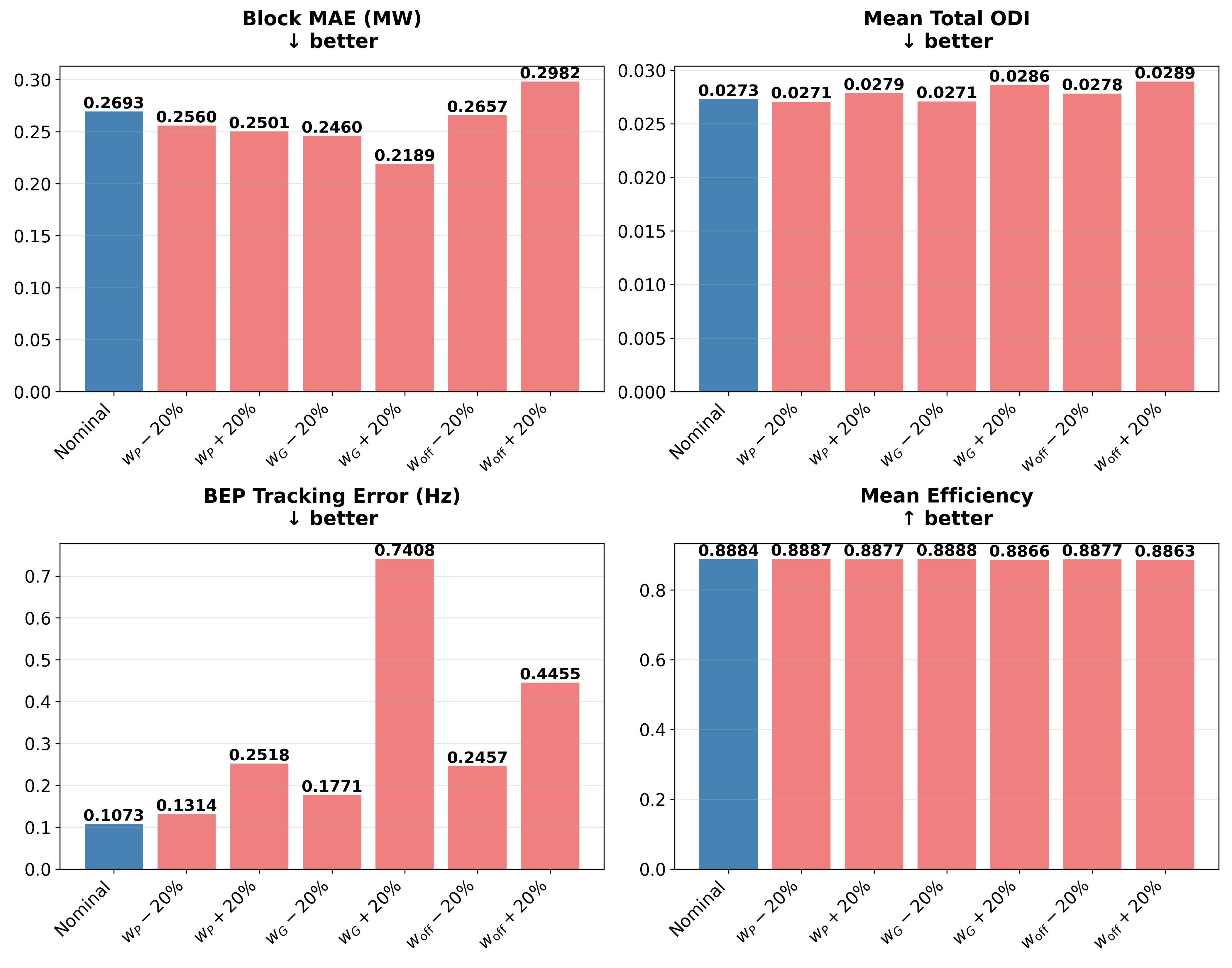}
}
\\[1mm]
\subfigure[Stress-scenario ODI sensitivity]{
\includegraphics[width=\columnwidth]{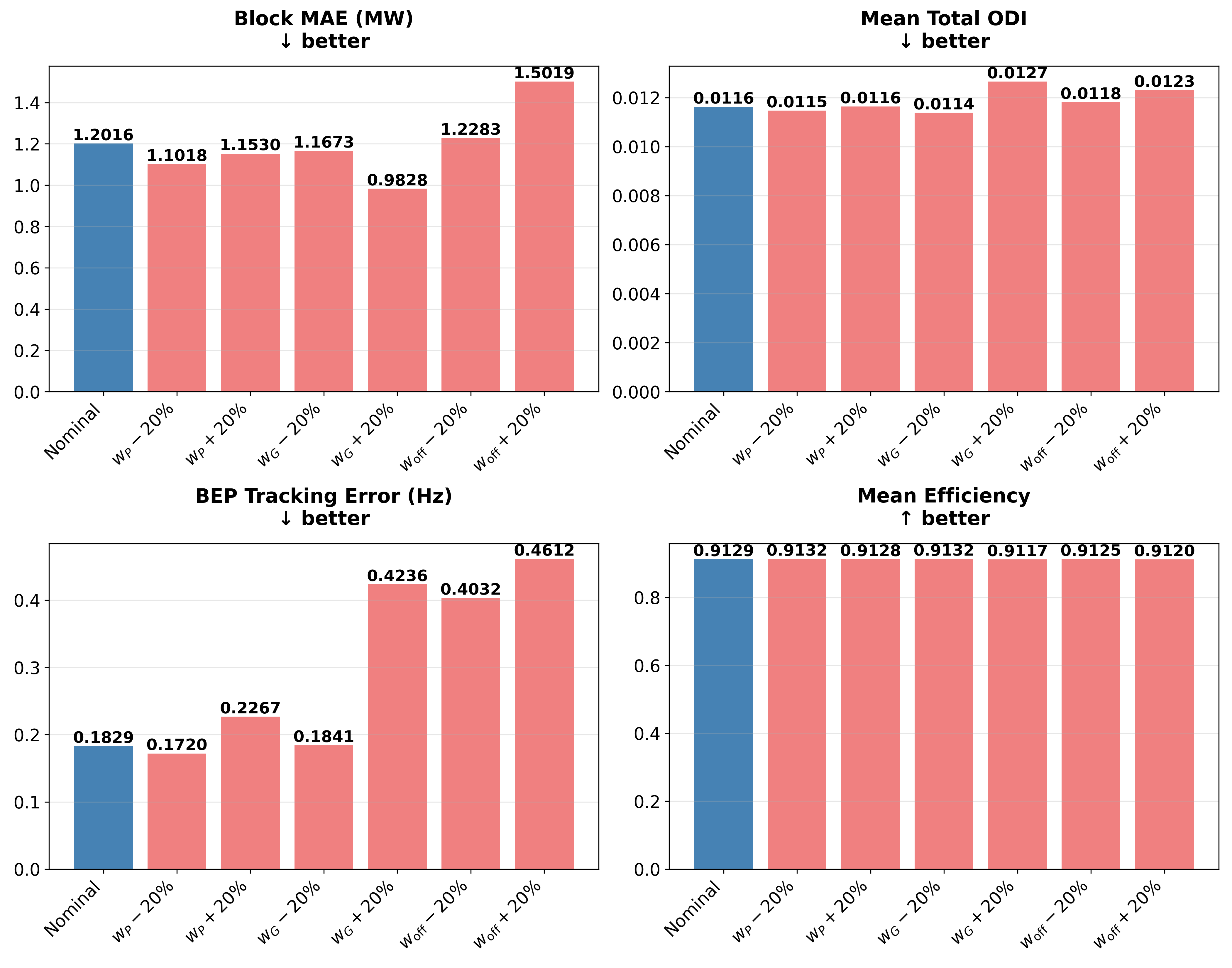}
}
\caption{Sensitivity of DA-RRL to $\pm 20\%$ variations in the ODI reward weights ($w_P$, $w_G$, $w_{\mathrm{off}}$) in the normal (a) and stress (b) scenarios.}
\label{fig:odi_sensitivity}
\end{figure}

In the second test, each ODI penalty weight ($w_P$, $w_G$, $w_{\mathrm{off}}$) is independently perturbed by $\pm20\%$ in the training reward, and the RL policy is retrained for each configuration. Figure~\ref{fig:odi_sensitivity} shows that total ODI is largely insensitive to these variations (approximately 0.027 in normal and 0.011 in stress). The most consequential perturbation is a $+20\%$ increase in $w_G$: by penalizing gate movement more heavily, it discourages the modulations needed to sustain power balance, which forces the rotor speed to drift and worsens BEP tracking error, particularly under stress. Changes to $w_P$ and $w_{\mathrm{off}}$ produce smaller fluctuations in MAE and BEP alignment without altering the relative ordering of methods. The baseline weight values thus represent a suitable operating point that balances the competing objectives.

%

\section{Conclusion}\label{sec:conclusion}
This paper proposed a degradation-aware residual control architecture for variable-speed pumped storage in pumping mode. The proposed architecture preserves a deterministic FF-PI block-tracking loop for certifiable 5-minute dispatch compliance, and augments it with a PPO-trained residual RL policy responsible solely for optimizing the rotor speed set-point. This structural separation bounds the worst-case speed command by construction, allowing the policy to pursue BEP tracking without interfering with power regulation.

The ODI synthesizes three physically interpretable fatigue drivers into a single composite reward signal. Embedding ODI and an efficiency bonus into the per-step reward enables DA-RRL to learn a speed trajectory that follows the demand-dependent BEP map while preserving practical tracking. Numerical evaluation yields four key findings: (i) DA-RRL reduces BEP error by about {\KBU 96\% (normal) and 95\% (stress)} relative to FF-PI; (ii) DA-RRL reduces total ODI by about {\KBU 73\% (normal) and 56\% (stress)} relative to FF-PI; (iii) DA-RRL achieves the lowest ODI among all compared methods in both scenarios while maintaining much lower MAE than MPC-D; and (iv) sensitivity analysis confirms stronger robustness of DA-RRL than FF-PI under degradation-model uncertainty.

Future work will extend the formulation to generation mode, investigate transfer across machines, and explore multi-unit coordination for VS-PSH plants sharing a common penstock.

%

\bibliography{references2}
\bibliographystyle{IEEEtran}

\end{document}